\documentclass[amssymb,nobibnotes,aps,prd,preprintnumbers]{revtex4}
 \usepackage{here}

\usepackage[dvipdfmx]{graphicx}
\usepackage{amssymb,amsfonts,amsmath,bm,color,multirow,cases,empheq,hyperref}
\usepackage{mathrsfs}

\usepackage{enumerate}
\usepackage{ulem}
\usepackage{afterpage}
\usepackage{ulem}
\usepackage{wrapfig}
\usepackage{ulem}

\newcommand{\del}[1]{ \partial_{#1} }

\begin{document}

\title{\Large 
Nonlinear dynamics driving the conversion of gravitational and electromagnetic waves in cylindrically symmetric spacetime
 }

\author{Takashi Mishima${}^1$\footnote{mishima.takashi@nihon-u.ac.jp} and Shinya Tomizawa${}^{2}$\footnote{tomizawa@toyota-ti.ac.jp}}
\vspace{2cm}
\affiliation{
${}^1$ Laboratory of Physics, College of Science and Technology, Nihon University,
Narashinodai, Funabashi, Chiba 274-8501, Japan\\
${}^2$ 
Mathematical Physics Laboratory, Toyota Technological Institute,
Hisakata 2-12-1, Nagoya 468-8511, Japan
}

\begin{abstract}
Using the ``composite harmonic mapping method," we construct exact solutions for cylindrically symmetric gravitational and electromagnetic waves within the Einstein-Maxwell system, focusing on the conversion dynamics between these types of waves. In this approach, we employs two types of geodesic surfaces in ${\rm H^{2}_{C}}$: (a) the complex line and (b) the totally real Lagrangian plane, applied to two different vacuum seed solutions: (i) a vacuum solution previously utilized in our studies and (ii) the solitonic vacuum solution constructed previously  by Economou and Tsoubelis.
We study three scenarios: case~(a) with seeds (i) and (ii), and case~(b) with seed~(ii).
In all cases (a) and (b), solutions demonstrate notable mode conversions near the symmetric axis.
In case~(a) with seed~(i) or seed~(ii),
we show that any change in the occupancy of the gravitational or electromagnetic mode relative to the C-energy near the axis always reverts to its initial state once the wave moves away from the axis.
Particularly in case~(b) with seed~(ii), nontrivial conversions occur even when the wave moves away from the axis. 
In this case, the amplification factors of electromagnetic modes range from an upper limit of approximately $2.4$ to a lower limit of about $0.4$, when comparing the contributions of electromagnetic mode to C-energy at past and future null infinities.
\end{abstract}
\preprint{TTI-MATHPHYS-29}

\maketitle

\section{Introduction}
\label{sec:1}

A series of gravitational wave detections since the first in 2015~\cite{LIGOScientific:2016aoc} (for more references, see the LIGO webpage:~\cite{LIGOScientific:List}.) have opened up a new era of discovery in gravitational physics~\cite{Barack:2018yly}.
In the future, these discoveries must provide us with clues to the secrets of gravitation in the universe.
Indeed, based on these detections of gravitational waves, tests of general relativity have been conducted~\cite{LIGOScientific:2016lio,LIGOScientific:2018dkp,LIGOScientific:2020tif}.
Contrary to expectations, no signs of new physics beyond Einstein's theory of gravity have been found so far, but future improvements in observational accuracy will likely lead to breakthroughs in new aspects of gravitational physics.
Therefore, since no promising modified theory of gravity has yet emerged in a clear form, we believe it is meaningful to explore the hidden aspects of the original theory of gravity, i.e., Einstein's theory of gravity.
Especially in the near future, fundamental research on nonlinear effects arising from the interaction of gravitational waves with waves of matter fields will become more important in order to reveal fundamental and unexpected aspects of gravity through observed gravitational waves.
As a first step, considering the Einstein-Maxwell system is simple and suitable for investigating such nonlinear effects.

\medskip
Our aim in this paper is to elucidate the fundamental and mathematical aspects of gravitational physics, particularly through the use of exact solutions to the Einstein-Maxwell equations.
While numerical general relativity offers a practical and versatile approach~\cite{book numerical relativity BS, book numerical relativity Shibata, Kyutoku:2021icp}, traditional methods based on exact solutions are subject to limitations due to necessary assumptions, such as spacetime symmetries.
However, with an appropriate exact solution, this method has provided deeper insights into the fundamentals of gravity from various angles~\cite{bicak:2000,book exact solution,book Exact Space-Times,book colliding waves,BronnikovSantosWang2020}.
Indeed, well-known instances like the Schwarzschild solution and the Einstein-Rosen waves have yielded significant discoveries about black holes and gravitational waves, respectively, through exact solution analysis.
Thus, this method will continue to have the potential to uncover new nonlinear effects in strong gravitational fields.
From the aforementioned perspective, we aim to elucidate novel and intriguing aspects of nonlinear effects in strong gravitational waves.
To this end, we have developed a new, straightforward, and compelling wave solution of the Maxwell-Einstein equations with cylindrical symmetry, employing a simple harmonic map method.
Our analysis of this solution revealed significant conversion between gravitational and electromagnetic waves under certain conditions~\cite{MT2022}.
Remarkably, this phenomenon can occur without the presence of intense background fields, such as the electromagnetic fields surrounding black holes, which typically act as a ``catalyst."
Thus, the substantial conversion observed is attributed to the nonlinear dynamics resulting from the interaction between the waves.
This result distinguishes our work from previous studies that examined similar conversion phenomena through perturbative approaches~\cite{Gelrach1974, OlsonUnruh1974, Matzner1976, CrispinoHiguchiOliveira2009,SaitoSodaYoshino2021,Cabral:2016klm,OuldElHadj:2021fqi, OuldElHadj:2023anu}, 
 and likely represents the first demonstration of large conversion phenomena driven entirely by ``nonlinearity."
In fact, a numerical relativistic approach had already been used to study the conversion phenomena~\cite{BarretoOliveiraRodrigues:2017}.
As mentioned above, however, the use of exact solutions can reveal new phenomena in the Einstein-Maxwell system, especially from a global perspective or in the case of extremely strong fields.

\medskip
In this paper, we primarily focus on the characteristic nonlinear effect in the Einstein-Maxwell system, specifically the conversion phenomena between gravitational waves and electromagnetic waves.
Here, we adopt a setup similar to our previous works on cylindrically symmetric gravitational waves~\cite{MishimaTomizawa2017}.
The background spacetime is regular and locally flat. A wave with a finite wave packet-like shape enters the vicinity of the axis from past null infinity, intensifies near the axis, reflects off the axis, and moves toward future null infinity.
In cylindrical symmetric spacetimes, we can introduce the cylindrical energy (i.e., so-called C-energy) and related quantities (i.e., density, flux) to measure how the conversion occurs. We will, therefore, use the occupancy that represents the extent of each mode's contributions to the C-energy density.
The occupancy is a local quantity that depends only on the radial coordinate $\rho$ and the time coordinate $t$.
As such, the occupancy as a function of $\rho$ changes over time as the waves move.
We expect that the occupancy near the axis drastically changes when the wave moves into the intense region of the fields near the axis.
In our previous paper, we noted that large conversions can occur near the axis.
Therefore, based on the previous result, we will treat such conversion phenomena in a more systematic manner.
However, the conversions occurring in the vicinity of the axis may disappear away from the axis; in fact, the solution considered in the previous paper is of this type.
Therefore, another interesting question arises~\cite{Harada}: if the incident and reflected waves are observed far enough off-axis, whether and to what extent non-trivial conversions (especially between gravitational and electromagnetic modes) occur, i.e., whether the conversions  that occur near the axis still remain after the waves move away from the axis.
In this second paper, we examine the above problem by extending the solution to soliton solutions.

\medskip
This paper is organized as follows:
In Section II, we first introduce the basic equations derived from the vacuum Einstein-Maxwell equations under the assumption of cylindrical symmetry.
Following the work in Ref.~\cite{Piran}, we define the amplitudes of cylindrical gravitational waves as useful tools for analysis. We then prepare useful quantities such as C-energy, among others.
In Section III, we explain the solution generation method adopted here and present the solutions derived by this method.
In Section IV, after analyzing the spacetime structures represented by the obtained solutions, we analyze the conversion phenomena between gravitational and electromagnetic waves.
In Section V, we provide a summary and discussion.

\section{Basic Equation and Fundamental Quantities}
\label{sec:2}

In this paper, we explore the interaction between gravitational and electromagnetic waves in cylindrically symmetric spacetime. Initially, we utilize the Kompaneets-Jordan-Ehlers form of the line element, as detailed in \cite{Kompaneets-Jordan-Ehlers} (refer to \cite{BronnikovSantosWang2020} for systematic treatments of cylindrical symmetric systems), along with a gauge potential. The line element and gauge potential are expressed as follows:
\begin{eqnarray}
ds^{2}
 &=& e^{2\psi}(dz-w d\phi)^2+\rho^2 e^{-2\psi}d\phi^2+e^{2(\gamma-\psi)}(-dt^2+d\rho^2),  
  \label{eq:KJE}  \\
{\bf A}
 &=& A_{\phi}d\phi + A_{z}dz,  
  \label{eq:gaugepot}
\end{eqnarray}
where  $\partial/\partial z$ and $\partial/\partial \phi$ are the Killing vectors corresponding to translational and rotational symmetries, respectively.  The metric functions $\psi$, $w$, $\gamma$, and the gauge potentials $A_{\phi}$, $A_{z}$ are functions only of the coordinates $t$ and $\rho$. 
We employ the geometric unit system where both the speed of light $c$ and the gravitational constant $G$ are set to 1.

\subsection{Basic equations}
By inserting Eqs.~(\ref{eq:KJE}) and (\ref{eq:gaugepot}) into the Einstein-Maxwell equations below
\begin{eqnarray}
R^{\mu}_{\ \nu}-\frac{1}{2}\delta^{\mu}_{\ \nu}R
&=& 8\pi T^{\mu}_{\ \nu}, \ \ \ 
 T^{\mu}_{\ \nu}=\frac{1}{4\pi}\left(
 F^{\mu\rho}F_{\nu\rho} - \frac{1}{4}\delta^{\mu}_{\ \nu}F^{\rho\delta} 
 F_{\rho\delta} \right) ,
  \label{eq:EinsteinEq}  \\
\nabla_{\mu}F^{\mu}_{\ \nu}
&=&0, 
  \label{eq:MaxwellEq}
\end{eqnarray}
and then algebraically manipulating these equations, we can derive the fundamental equations governing the fields $\psi$, $w$, $A_{\phi}$, and $A_{z}$, as follows:
\begin{eqnarray}
\nabla^{2}\psi
&=& 
    -\frac{1}{2\rho^2}e^{4\psi} \left[
        (\del{t}w)^2 - (\del{\rho}w)^2 \right] 
    + e^{-2\psi}\left[ (\del{t}A_{z})^2 - (\del{\rho}A_{z})^2 \right]  \nonumber  \\
 && - \frac{1}{\rho^2}e^{2\psi}\left[ 
     (\del{t}A_{\phi} + w\del{t}A_{z} )^2 -  (\del{\rho}A_{\phi} + w\del{\rho}A_{z} )^2 \right],   
  \label{eq:EinsteinEq1}
\end{eqnarray}
\begin{eqnarray}
-\del{t}^{2}w - \frac{1}{\rho}\del{\rho}w + \del{\rho}^{2}w
&=& 
       4(\del{t}\psi \del{t}w - \del{\rho}\psi\del{\rho}w)     \nonumber     \\
&&   - 4 e^{-2\psi}
       \left[
         \del{t}A_{z} (\del{t}A_{\phi} + w \del{t}A_{z} ) - \del{\rho}A_{z} (\del{\rho}A_{\phi} + w \del{\rho}A_{z} )
       \right],
  \label{eq:EinsteinEq2}
\end{eqnarray}
\begin{eqnarray}
-\del{t}^{2}A_{\phi} - \frac{1}{\rho}\del{\rho}A_{\phi} + \del{\rho}^{2}A_{\phi}
&=&
 w( \del{t}^{2}A_{z} + \frac{1}{\rho}\del{\rho}A_{z} - \del{\rho}^{2}A_{z})+(\del{t}w \del{t}A_{z} - \del{\rho}w\del{\rho}A_{z})        \nonumber \\
    &&   
    +2 \del{t}\psi (\del{t}A_{\phi} + w \del{t}A_{z}) - 2 \del{\rho}\psi (\del{\rho}A_{\phi} + w \del{\rho}A_{z}), 
  \label{eq:MaxwellEq1} 
\end{eqnarray}
\begin{eqnarray}
\nabla^{2}A_{z}
&=&  - 2\del{t}\psi \del{t}A_{z}+ 2\del{\rho}\psi \del{\rho}A_{z}  \nonumber  \\
&&  + \frac{1}{\rho^2}e^{4\psi} \bigl[ 
                                                    \del{t}w ( \del{t}A_{\phi}+w\del{t}A_{z}) 
                                                     -\del{\rho}w (\del{\rho}A_{\phi} +w \del{\rho}A_{z} \bigr) 
                                                     \bigr].
  \label{eq:MaxwellEq2} 
\end{eqnarray}
In this framework, $\nabla$ denotes the covariant derivative in the three-dimensional Minkowski space $M^{(1,2)}$. The line element in cylindrical coordinates is expressed as

\begin{eqnarray}
ds^2 =h_{\mu\nu}dx^{\mu}dx^{\nu}=
-dt^2 + d\rho^2 + \rho^2 d\phi^2,
  \label{eq:line-element1}
\end{eqnarray}
where the Greek subscripts $\mu$ and $\nu$ sequentially take the indices corresponding to the coordinates $(t, \rho, \phi)$.
Moreover, recognizing that Eqs.~(\ref{eq:EinsteinEq2}) and (\ref{eq:MaxwellEq1}) correspond to the following integrability conditions:
\begin{eqnarray}
\del{t}\left[\frac{e^{4\psi}}{\rho}\del{t}w
         -4\frac{e^{2\psi} }{\rho}A_{z} \left(\del{t}A_{\phi} + w \del{t}A_{z} \right)  \right]
&=&   
\del{\rho}\left[\frac{e^{4\psi}}{\rho}\del{\rho}w
         -4\frac{e^{2\psi} }{\rho}A_{z} \left(\del{\rho}A_{\phi} + w \del{\rho}A_{z} \right)  \right],     \\
\del{t} \Bigl[ \frac{1}{\rho}e^{2\psi} \bigl( \del{t}A_{\phi} + w\del{t}A_{z} \bigr) \Bigr]
&=&\del{\rho} \Bigl[ \frac{1}{\rho}e^{2\psi} \bigl( \del{\rho}A_{\phi} + w\del{\rho}A_{z} \bigr) \Bigr], 
  \label{eq:MaxwellEq4} 
\end{eqnarray}
we can introduce potential functions $\chi$ and $\Phi$, defined by the respective equations:
\begin{eqnarray}
(\, \del{t} \chi,\ \del{\rho} \chi \, )
&=&\left( \frac{1}{\rho} e^{2\psi} \bigl( \del{\rho}A_{\phi} + w\del{\rho}A_{z} ),\    
\frac{1}{\rho} e^{2\psi}( \del{t}A_{\phi} + w\del{t}A_{z} )  \right),  
  \label{eq:potdef1}   \\
(\del{t}\Phi,\, \del{\rho}\Phi )
&=&
 \left(\, \frac{1}{\rho}e^{4\psi}\del{\rho}w  -2(A_{z} \del{t}\chi - \chi \del{t}A_{z} ) ,\ 
  \frac{1}{\rho}e^{4\psi}\del{t}w -2(A_{z} \del{\rho}\chi - \chi \del{\rho}A_{z} ) \, \right) .  
  \label{eq:potdef2}
\end{eqnarray}
Thus, the basic equations (\ref{eq:EinsteinEq1}), (\ref{eq:EinsteinEq2}), (\ref{eq:MaxwellEq1}), and (\ref{eq:MaxwellEq2}) can be reformulated in terms of the variables $\psi$, $\Phi$, $\chi$, and $A_{z}$:
\begin{eqnarray}
\nabla^2\psi
&=& 
    \frac{1}{2}e^{-4\psi} \left[
    (\del{t}\Phi + 2A_{z}\del{t}\chi - 2\chi\del{t}A_{z})^2-
      (\del{\rho}\Phi + 2A_{z}\del{\rho}\chi - 2\chi\del{\rho}A_{z})^2
                               \right]                                        \nonumber  \\
  &&  + e^{-2\psi}\left[ (\del{t}A_{z})^2 - (\del{\rho}A_{z})^2 \right]  
        + e^{-2\psi}\left[ (\del{t}\chi)^2 - (\del{\rho}\chi)^2 \right],       
  \label{eq:Ernstpart1}  \\
\nabla^2\Phi
&=& 
       4\left[-\del{t}\psi( \del{t}\Phi +A_{z}\del{t}\chi - \chi\del{t}A_{z} ) 
                + \del{\rho}\psi( \del{\rho}\Phi +A_{z}\del{\rho}\chi - \chi\del{\rho}A_{z} )  \right]  \nonumber  \\
 &&     +e^{-2\psi}\left[ -(\del{t}\Phi +2A_{z}\del{t}\chi - 2\chi\del{t}A_{z} )(\del{t}A_{z}^2 +\del{t}\chi^2) \right. \nonumber \\
 &&     \qquad\quad \left.        + (\del{\rho}\Phi +2A_{z}\del{\rho}\chi - 2\chi\del{\rho}A_{z} )(\del{\rho}A_{z}^2 +\del{\rho}\chi^2)
                            \right],             
  \label{eq:Ernstpart2}    \\
\nabla^2A_{z}
&=&  - 2\del{t}\psi \del{t}A_{z}+ 2\del{\rho}\psi \del{\rho}A_{z}      \nonumber  \\
&&  + e^{-2\psi} \bigl[ 
                                 \del{\rho}\chi ( \del{\rho}\Phi + 2A_{z}\del{\rho}\chi - 2\chi\del{\rho}A_{z} ) 
                                                     -\del{t}\chi ( \del{t}\Phi + 2A_{z}\del{t}\chi - 2\chi\del{t}A_{z} \bigr) 
                                                     \bigr],     
  \label{eq:Ernstpart3}    \\
\nabla^2\chi
&=& 
   -2\del{t}\psi \del{t}\chi + 2\del{\rho}\psi \del{\rho}\chi  
     + e^{-2\psi}\del{t}A_{z} \left[\del{t}\Phi + 2(A_{z}\del{t}\chi - \chi\del{t}A_{z}) \right]   \nonumber    \\
&& -e^{-2\psi}\del{\rho}A_{z} \left[\del{\rho}\Phi + 2(A_{z}\del{\rho}\chi - \chi\del{\rho}A_{z}) \right]. 
  \label{eq:Ernstpart4}
\end{eqnarray}
The wave equations  (\ref{eq:Ernstpart2}) and (\ref{eq:Ernstpart4}) have been derived from the compatibility of 
the definitions (\ref{eq:potdef1}) and (\ref{eq:potdef2}), i.e. 
$\del{t}\del{\rho}w=\del{\rho}\del{t}w$ and $\del{t}\del{\rho}A_{\phi}=\del{\rho}\del{t}A_{\phi}$, respectively.
Next, we introduce two complex potentials ${E}$ and ${F}$ as follows:
\begin{eqnarray}
{E}:= e^{2\psi} +|{F}|^2- i \Phi,\ \ \ {F}:= A_{z} + i\chi.  
  \label{eq:Ernst pot}
\end{eqnarray}
Using these potentials, the set of wave equations (\ref{eq:Ernstpart1})--(\ref{eq:Ernstpart4}) can be succinctly expressed in the form of the cylindrically symmetric version of the Ernst equations \cite{Ernst68}:
\begin{eqnarray}
&&({\rm Re}[E ]-{F} \bar{{F}} )\, \nabla^2E= (\nabla E
   - 2\bar{F}\nabla{F} )\cdot \nabla{E},
\label{eq:Ernst11}  \\
&&({\rm Re}[{E} ]-{F} \bar{{F}} )\, \nabla^2{F}= (\nabla{E}
   - 2\bar{F}\nabla{F} )\cdot \nabla{F},
\label{eq:Ernst12}  
\end{eqnarray}
where the dot $(\cdot)$ means the scalar product defined with the metric~(\ref{eq:line-element1}).

\subsection{Construction of metric and electromagnetic fields} 

The methodology employed in this study to construct a new set of metric and electromagnetic fields is as follows: Initially, we solve Eqs.~(\ref{eq:Ernst11}) and (\ref{eq:Ernst12}) to find their solutions. Next, utilizing the definition (\ref{eq:Ernst pot}), we algebraically determine the quantities  $e^{2\psi}$, $\Phi$, ${\chi}$ and $A_{z}$.
Subsequently, employing the relations~(\ref{eq:potdef1}) and (\ref{eq:potdef2}), we derive the metric function  $w$ and the gauge field component~$A_{\phi}$ through integration. 
The initial step is crucial due to the nonlinearity of the fundamental equations. 
Detailed elaboration of the first step will be provided later.

\medskip
The remaining metric function $\gamma$ can be determined by the following equations:
\begin{eqnarray}
\del{\rho}\gamma
&=& \rho\left[ (\del{t}\psi)^2+  (\del{\rho}\psi)^2 \right]
   +\frac{1}{4\rho}e^{4\psi}\left[
     \left(\del{t}w \right)^2+ \left(\del{\rho} w \right)^2 \right]  \nonumber  \\
&&   +\rho e^{-2\psi}\left[
     \left(\del{t}A_{z} \right)^2+ \left(\del{\rho} A_{z} \right)^2 \right]
      +\frac{1}{\rho}e^{2\psi}\left[
     \left(\del{t}A_{\phi}+w\del{t}A_{z} \right)^2 
     + \left(\del{\rho}A_{\phi}+w\del{\rho}A_{z} \right)^2 \right] ,
\label{eq:gamma1}   \\ 
\del{t}\gamma
&=& 2\rho\del{t}\psi \del{\rho}\psi + \frac{1}{2\rho}e^{4\psi}\del{t}w \del{\rho}w  \nonumber  \\
&&      + 2\rho e^{-2\psi} \del{t}A_{z} \del{\rho}A_{z}
        +\frac{2}{\rho}e^{2\psi}\left(\del{t}A_{\phi}+w\del{t}A_{z} \right)
      \left(\del{\rho}A_{\phi} + w\del{t\rho}A_{z} \right).
\label{eq:gamma2} 
\end{eqnarray}
These equations are derived from the $(t,t)$ and $(t,\rho)$ components of the Einstein equation (\ref{eq:EinsteinEq}). Therefore, after calculating the other quantities, $\gamma$ can be obtained by integrating Eqs.~(\ref{eq:gamma1}) and (\ref{eq:gamma2}). The integrability condition for these equations is ensured by other equations in the system.
It is important to note that the sum of the third and fourth terms on the right-hand side of Eqs.~(\ref{eq:gamma1}) and (\ref{eq:gamma2}) corresponds to the components of the electromagnetic stress-energy tensor, ${}^{(\rm em)} T_{tt}$ or ${}^{(\rm em)}T_{t\rho}$, scaled by $\rho/8\pi$, respectively.

\subsection{Amplitudes, C-energy, and related quantities} 

We introduce key quantities for analyzing the physical behavior of cylindrically symmetric gravitational systems, particularly concerning mode conversion phenomena. 
Following the methodology of Piran, Safier, and Stark \cite{Piran}, and extending the Einstein-Maxwell system as per \cite{MT2022}, we introduce the `amplitudes' as follows:
\begin{eqnarray}
{\cal A}_{+} &:=& 2\ \del{v}\psi ,\ \ 
{\cal A}_{\times} := \frac{1}{\rho}e^{2\psi}\del{v}w= e^{-2\psi} \bigl[ \del{v}\Phi +2(A_{z}\del{v}\chi - \chi\del{v}A_{z} ) \bigr], \nonumber \\
{\cal A}_{z} &:=& 2e^{-\psi} \del{v}A_{z}, \ \ \ 
{\cal A}_{\phi} := \frac{2}{\rho}e^{\psi}\left(\del{v}A_{\phi}+w\del{v}A_{z} \right) = 2e^{-\psi} \del{v}\chi, \nonumber \\
{\cal B}_{+} &:=& 2\ \del{u}\psi , \ \ 
{\cal B}_{\times} := \frac{1}{\rho}e^{2\psi}\del{u}w = -e^{-2\psi} \bigl[ \del{u}\Phi +2(A_{z}\del{u}\chi - \chi\del{u}A_{z} ) \bigr] , \nonumber \\
{\cal B}_{z} &:=& 2e^{-\psi} \del{u}A_{z},\ \ \ 
{\cal B}_{\phi}:= \frac{2}{\rho}e^{\psi}\left(\del{u}A_{\phi}+w\del{u}A_{z} \right) = -2e^{-\psi} \del{u}\chi, 
\label{eq:amplitudes} 
\end{eqnarray}
where ${\cal A}_{(\cdot)}$ and ${\cal B}_{(\cdot)}$ represent the `ingoing' and `outgoing' amplitudes, respectively, with the subscripts designating the corresponding mode. Furthermore, the null coordinates $u = (t - \rho)/2$ and $v = (t + \rho)/2$ are used for this formulation.
The fundamental equations for the amplitudes can be derived from the field equations for $\psi$, $w$, $A_{\phi}$, and $A_{z}$ (Eqs.~(\ref{eq:EinsteinEq1})--(\ref{eq:MaxwellEq2})) as follows:
\begin{eqnarray}
\del{u} {\cal A}_{+}
&=& \frac{1}{2\rho}( {\cal A}_{+} - {\cal B}_{+} ) + {\cal A}_{\times} {\cal B}_{\times} 
-\frac{1}{2}( {\cal A}_{z}{\cal B}_{z} - {\cal A}_{\phi}{\cal B}_{\phi} ),   \nonumber  \\
\del{v} {\cal B}_{+} 
&=& \frac{1}{2\rho}( {\cal A}_{+} - {\cal B}_{+} ) + {\cal A}_{\times} {\cal B}_{\times} 
-\frac{1}{2}( {\cal A}_{z}{\cal B}_{z} - {\cal A}_{\phi}{\cal B}_{\phi} ),  \nonumber  \\
\del{u} {\cal A}_{\times} 
&=& \frac{1}{2\rho}( {\cal A}_{\times} + {\cal B}_{\times} ) - {\cal A}_{+} {\cal B}_{\times} 
+\frac{1}{2}( {\cal A}_{\phi}{\cal B}_{z} + {\cal A}_{z}{\cal B}_{\phi} ),    \nonumber  \\
\del{v} {\cal B}_{\times}
&=& -\frac{1}{2\rho}( {\cal A}_{\times} + {\cal B}_{\times} ) - {\cal A}_{\times} {\cal B}_{+} 
+\frac{1}{2}( {\cal A}_{\phi}{\cal B}_{z} + {\cal A}_{z}{\cal B}_{\phi} ),    \nonumber  \\
\del{u} {\cal A}_{\phi} 
&=& \frac{1}{2\rho}( {\cal A}_{\phi} + {\cal B}_{\phi} ) + \frac{1}{2}( {\cal A}_{z}{\cal B}_{\times} 
- {\cal A}_{\times}{\cal B}_{z} -{\cal A}_{+}{\cal B}_{\phi}),   \nonumber  \\
\del{v} {\cal B}_{\phi} 
&=& -\frac{1}{2\rho}( {\cal A}_{\phi} + {\cal B}_{\phi} ) - \frac{1}{2}( {\cal A}_{z}{\cal B}_{\times} 
- {\cal A}_{\times}{\cal B}_{z} +{\cal A}_{\phi}{\cal B}_{+}),      \nonumber  \\
\del{u} {\cal A}_{z} 
&=& \frac{1}{2\rho}( {\cal A}_{z} - {\cal B}_{z} ) - \frac{1}{2}( {\cal A}_{\phi}{\cal B}_{\times} 
+ {\cal A}_{\times}{\cal B}_{\phi} -{\cal A}_{+}{\cal B}_{z}),   \nonumber  \\
\del{v} {\cal B}_{z} 
&=& \frac{1}{2\rho}( {\cal A}_{z} - {\cal B}_{z} ) - \frac{1}{2}( {\cal A}_{\phi}{\cal B}_{\times} 
+ {\cal A}_{\times}{\cal B}_{\phi} -{\cal A}_{z}{\cal B}_{+}).  
\label{eq:AmplitudesEq} 
\end{eqnarray}
These equations are a generalization of the basic equations presented in the previous work~\cite{Piran}.

\medskip
Furthermore, in analogy with the concept of C-energy in cylindrically symmetric vacuum spacetimes, we introduce local quantities related to the C-energy \cite{Thorn65}, denoted as ${\cal E} := \partial_{\rho}\gamma$ and ${\cal F} := \partial_{t}\gamma$.
From Eqs.~(\ref{eq:gamma1}) and (\ref{eq:gamma2}), the quantities ${\cal E}$ and ${\cal F}$ can be expressed in terms of the amplitudes from Eq.~(\ref{eq:amplitudes}) as follows:
\begin{eqnarray}
{\cal E}_{}&=& 
 \frac{\rho}{8} 
\left( {\cal A}_{+}^2 + {\cal B}_{+}^2 + {\cal A}_{\times}^2 + {\cal B}_{\times}^2 
+ {\cal A}_{z}^2 + {\cal B}_{z}^2 + {\cal A}_{\phi}^2 + {\cal B}_{\phi}^2 \right),    
\label{eq:CenergyFlux-1}    \\
{\cal F}_{}&=& 
 \frac{\rho}{8} 
\left( {\cal A}_{+}^2 - {\cal B}_{+}^2 + {\cal A}_{\times}^2 - {\cal B}_{\times}^2 
+ {\cal A}_{z}^2 - {\cal B}_{z}^2 + {\cal A}_{\phi}^2 - {\cal B}_{\phi}^2 \right).
\label{eq:CenergyFlux-2} 
\end{eqnarray}
Using Eqs.~(\ref{eq:AmplitudesEq}), we can verify the local conservation law $-\partial_{t}{\cal E} + \partial_{\rho}{\cal F} = 0$ through some algebraic manipulations. 
This conservation law allows us to define a type of total energy, specifically the C-energy, per unit length along the coordinate $z$. This energy is contained within a cylindrical region of radius $\rho_0$ at a given time:
\begin{eqnarray}
E_{}(t,\rho_0)=\int_0^{\rho_0}{\cal E}d\rho=\gamma(t,\rho_0)-\gamma(t,0).
\label{eq:Cenergy-1} 
\end{eqnarray}
In Eq.~(\ref{eq:Cenergy-1}), the term $\gamma(t,0)$ vanishes when the physical quantities are regular at the axis, ensuring the physical relevance of the defined C-energy.

\medskip
For the remainder of the discussion, we focus on the case where $\gamma(t,0) = 0$, which aligns with our interest in scenarios devoid of singular field sources on the axis of symmetry. In such instances, it is reasonable to anticipate that the C-energy for a suitably regular, packet-like wave, denoted as $E(t, \infty)$, remains finite constant over time.
To further dissect the contribution of each mode to the C-energy, we define the following quantities:
\begin{eqnarray}
E_{ I}(t,\rho_0)=\int_0^{\rho_0}{\cal E}_{ I}d\rho,
\label{eq:Cenergy-2} 
\end{eqnarray}
where ${\cal E}_{ I}$ represents each mode contribution to the C-energy density~(\ref{eq:CenergyFlux-1}):
\begin{eqnarray}
{\cal E}_{ I}:= \frac{\rho}{8}( {\cal A}_{I}^2 + {\cal B}_{I}^2 ), \ \ \ ({ I} = +,\ \times,\ z,\ \phi ).
\label{eq:Cenergy-3} 
\end{eqnarray}

\section{Construction of the solutions}\label{sec:3}
To investigate the conversion phenomena between gravitational and electromagnetic fields as nonlinear effects, it is essential to construct suitable solutions that represent these phenomena. 
A standard approach for deriving such solutions is the composite harmonic mapping method, which is both straightforward  and practical for our objectives due to its operational simplicity.

This method has already been partially utilized in our previous work~\cite{MT2022}.
In this section, we will review the generation method and its extensions. The method will be applied to the seed used in the previous work~\cite{MT2022} and the soliton solution explored by Economou and Tsubelis~\cite{Economou}. A detailed explanation of the composite harmonic mapping method is provided in Appendix \ref{appendix:A}.

\subsection{Preparation to generate the harmonic map}
\subsubsection{\bf Overview}

To facilitate the generation of solutions from a clearer perspective, we transform the original Ernst equations (\ref{eq:Ernst11}) and (\ref{eq:Ernst12}) into an alternative set of Ernst equations, 
as per Ernst~\cite{Ernst68}, Kinnersley~\cite{Kinnersley73}, and Mazur~\cite{Mazur:1983vi}. 
This transformation helps to elucidate the structure of the equations and streamline the process of finding solutions that encapsulate the nonlinear interaction between gravitational and electromagnetic fields.
The transformed Ernst equations are as follows:
\begin{eqnarray}
(\xi \bar{\xi}+\eta \bar{\eta}-1 )\, \nabla^2{\xi}
&=& 2(\bar{\xi} \nabla{\xi} + \bar{\eta} \nabla{\eta} )\cdot \nabla{\xi}, 
\label{eq:Ernst21} \\
(\xi \bar{\xi}+\eta \bar{\eta}-1 )\, \nabla^2{\eta}
&=& 2(\bar{\xi} \nabla{\xi} + \bar{\eta} \nabla{\eta} )\cdot \nabla{\eta},
\label{eq:Ernst22}
\end{eqnarray}
where the complex quantities $\xi$ and $\eta$ are related to the potentials $E$ and $F$ defined by (\ref{eq:Ernst pot}), as follows 
\begin{eqnarray}
\xi= \frac{E-1}{E+1},\ \ \ \ \eta= \frac{2F}{E+1},\ \ \ (\xi \bar{\xi}+\eta \bar{\eta}<1).
  \label{eq:Relation1}
\end{eqnarray}
Inversely, the potentials $E$ and $F$ can be expressed in terms of $\xi$ and $\eta$:
\begin{eqnarray}
E= \frac{1+\xi}{1-\xi},\ \ \ \ F= \frac{\eta}{1-\xi}.
  \label{eq:Relation2}
\end{eqnarray}
These relationships and equations form the basis for analyzing and generating solutions to represent the interaction between gravitational and electromagnetic fields in the context of Ernst's formalism.

\medskip
It is noticed that the equations (\ref{eq:Ernst21}) and (\ref{eq:Ernst22}) can be derived as Euler-Lagrange equations from the following Lagrangian:
\begin{eqnarray}
{\cal L}= \frac{\rho}{2} \frac{ \nabla{\xi}\cdot \nabla \bar{\xi}
 + \nabla{\eta}\cdot \nabla \bar{\eta}
-(\xi\nabla \eta- \eta\nabla\xi)\cdot 
(\bar{\xi}\nabla \bar{\eta}- \bar{\eta}\nabla\bar{\xi} ) }
{(\xi \bar{\xi}+\eta \bar{\eta}-1 )^2}.
  \label{eq:Lagrangian}
\end{eqnarray}
This formulation implies that equations (\ref{eq:Ernst21}) and (\ref{eq:Ernst22}) can be viewed as determining the harmonic map from the base space $M^{(1,2)}$ into the target space ${\rm H^2_C}$, which is the ball model of complex two-dimensional hyperbolic space, as described in the literature on complex hyperbolic geometry ~\cite{book Complex Hyperbolic Geometry, Parker2010}.
The line element of the target space, extracted from the Lagrangian~(\ref{eq:Lagrangian}), is expressed as
\begin{eqnarray}
d\ell^2=G_{AB}dz^{A}dz^{B}
  =\frac{ d\xi d\bar{\xi} + d\eta d\bar{\eta}
   -(\xi d\eta- \eta d\xi)(\bar{\xi} d\bar{\eta}- \bar{\eta} d\bar{\xi} ) }
   {(\xi \bar{\xi}+\eta \bar{\eta}-1 )^2},
  \label{eq:line-element2}
\end{eqnarray}
where the coordinates $z^{A}, z^{B}$, etc., correspond to $(\xi, \eta, \bar{\xi}, \bar{\eta})$.
The metric form associated with the line element (\ref{eq:line-element2}) is represented in matrix form as follows:
\begin{eqnarray}
\left( G_{AB} \right)= \frac{1}{2(\xi \bar{\xi}+\eta\bar{\eta} -1 )^2 }
\left(\begin{array}{@{}c|c@{}}
\mbox{\Large 0} &
\begin{matrix}
1-\eta\bar{\eta} & \eta\bar{\xi} \\
\xi\bar{\eta}       & 1-\xi\bar{\xi} 
\end{matrix}  \\
\hline
\begin{matrix}
1-\eta\bar{\eta} & \xi\bar{\eta} \\
\eta\bar{\xi}      & 1-\xi\bar{\xi} 
\end{matrix}   &
{\mbox{\Large 0}}
\end{array}\right).
\end{eqnarray}
This metric representation delineates the geometric structure of the target space. 

\medskip
The procedure outlined in Appendix \ref{appendix:A} can be summarized as follows: First, identify an embedding map to create a totally geodesic subspace within the target space, denoted as ${\rm H^2_C}$. Then, establish a harmonic map from the base space to this subspace, and finally, integrate both mappings.
Mathematically, it is established that ${\rm H^2_C}$ possesses only two isometrically distinct classes of totally geodesic subspaces: (a) the complex planes, corresponding to the Poincare' disc model (hereinafter referred to as ${\rm H^1_C}$), and (b) the totally real Lagrangian planes, corresponding to the Klein disc model (abbreviated as ${\rm K_d}$)~\cite{book Complex Hyperbolic Geometry, Parker2010}.
These subspaces, represented as two-dimensional real (or one-dimensional complex) hyperbolic surfaces, exhibit distinct Gaussian curvatures, rendering them non-isometric. Once a totally geodesic surface is chosen, the next step involves identifying a suitable harmonic map from the base space to this hyperbolic surface (either ${\rm H^1_C}$ or ${\rm K_d}$), serving as an intermediate target space. 
The final step essentially involves solving the vacuum Ernst equation.

\subsubsection{\bf Case~(a): complex plane}
First, as an illustrative example of a totally geodesic embedding map in the coordinates of ${\rm H^2_C}$, denoted as $(\xi, \eta)$, we consider the surface defined by
\begin{eqnarray}
(\xi, \eta) = (\cos 2\theta z, \sin 2\theta z),
\label{eq:case-a}
\end{eqnarray}
where $z$ is viewed as a complex coordinate that characterizes the subspace ${\rm H^1_C}$, and the parameter $\theta$ differentiates between various subspaces.
We can verify that the map (\ref{eq:case-a}) fulfills the second of Eqs.~(\ref{eq:TGC}), aligning with case~(a) previously discussed. The corresponding line element is given by
\begin{eqnarray}
d\ell^2 = \frac{ dz d\bar{z} }{(z \bar{z} - 1 )^2},
\label{eq:line-element3}
\end{eqnarray}
which represents the line element of the Poincar\'{e} disc model.
A similar ansatz to Eq.~(\ref{eq:case-a}) has been utilized in different contexts, such as analyzing colliding plane waves, as noted in Halilsoy's work~\cite{Halilsoy:1989ji}. 
Our previous research~\cite{MT2022} is based on this case.
Furthermore, this includes the particular case $\theta = \pi/2$, which was treated in the study by Xanthopoulos~\cite{Xanthopoulos:1987ha}. 
According to the aforementioned procedure, once a solution $\xi_v(x)$ of the vacuum Ernst equation is obtained:
\begin{eqnarray}
(\xi_{v} \bar{\xi}_{v}-1 )\, \nabla^2{\xi}_{v}
&=& 2\bar{\xi}_{v} \nabla{\xi}_{v} \cdot \nabla{\xi}_{v},
\label{eq:VacuumErnst}
\end{eqnarray}
by substituting the coordinate $z$ with $\xi_v$, we obtain the desired harmonic map as follows:
\begin{eqnarray}
(\xi(x), \eta(x)) = (\,\cos2\theta\, \xi_{v}(x),\sin2\theta\, \xi_{v}(x)\,).
  \label{eq:map(1)}
\end{eqnarray}

\medskip
Next, we represent $\xi_v$ as $\xi_1 + i \xi_2$, where $\xi_1$ and $\xi_2$ are the real and imaginary parts of $\xi_v$, respectively. Using (\ref{eq:Relation2}), we derive the following expressions for $E$ and $F$:
\begin{eqnarray}
E&=&\frac{1 - (\xi_{1}^2 + \xi_{2}^2) \cos^2 2\theta}
  {1 - 2 \xi_1 \cos 2\theta + (\xi_{1}^2 + \xi_{2}^2) \cos^2 2\theta }
  + i \frac{2 \xi_{2} \cos 2\theta}
      {1 - 2 \xi_1 \cos 2\theta + (\xi_{1}^2 + \xi_{2}^2) \cos^2 2\theta} , \\
F&=& \frac{[\xi_{1} - (\xi_{1}^2 + \xi_{2}^2) \cos 2\theta] \sin 2\theta } 
         {1 - 2 \xi_1 \cos 2\theta + (\xi_{1}^2 + \xi_{2}^2) \cos^2 2\theta}
   +i \frac{ \xi_{2} \sin 2\theta }{1 - 2 \xi_1 \cos 2\theta + (\xi_{1}^2 + \xi_{2}^2) \cos^2 2\theta }.
\end{eqnarray}
Following the definition (\ref{eq:Ernst pot}), the expressions for the metric function $\psi$, the gravitational twist potential $\Phi$, the gauge field component $A_{z}$, and the dual potential $\chi$ associated with $A_{\phi}$ are derived as follows:
\begin{eqnarray}
e^{2\psi} &=& \frac{1 - \xi_{1}^2 - \xi_{2}^2}
       {1 - 2 \xi_1 \cos 2\theta + (\xi_{1}^2 + \xi_{2}^2) \cos^2 2\theta},  \nonumber \\
\Phi &=& -\frac{2 \xi_{2} \cos 2\theta }
       {1 - 2 \xi_1 \cos 2\theta + (\xi_{1}^2 + \xi_{2}^2) \cos^2 2\theta},  \nonumber \\
A_{z} &=& \frac{[\xi_{1} - (\xi_{1}^2 + \xi_{2}^2) \cos 2\theta ] \sin 2\theta } 
       {1 - 2 \xi_1 \cos 2\theta + (\xi_{1}^2 + \xi_{2}^2) \cos^2 2\theta}, \nonumber \\
\chi &=& \frac{ \xi_{2} \sin 2\theta }
       {1 - 2 \xi_1 \cos 2\theta + (\xi_{1}^2 + \xi_{2}^2) \cos^2 2\theta}.
  \label{eq:setofsolution-1}
\end{eqnarray}
These equations explicitly define the fields in terms of $\xi_1$, $\xi_2$, and the angle $\theta$, encapsulating the relationship between the geometrical structure of the solution and the physical fields.

\subsubsection{\bf Case~(b): totally real Lagrangian plane}

The totally real Lagrangian plane $(\xi, \eta) = (v_{1}, v_{2})$, where $v_{1}$ and $v_{2}$ are real coordinates, forms a totally geodesic surface in ${\rm H^2_C}$. 
This surface constitutes a real two-dimensional hyperbolic surface, represented as the Klein disc model. New harmonic maps can be generated from the solutions of the vacuum Ernst equation by employing the coordinate transformation between the Klein disc model $(v_{1}, v_{2})$ and the Poincar\'{e} disc model $z = z_{1} + iz_{2}$, given by:
\begin{eqnarray}
v_{1} +i v_{2}
= \frac{ 2z }{ 1+|z|^2 },
\end{eqnarray}
or equivalently,
\begin{eqnarray}
v_{1} &=& \frac{ 2z_{1} }{ 1+z_{1}^2+z_{2}^2 } ,\ \ \ 
v_{2} = \frac{ 2z_{2} }{ 1+z_{1}^2+z_{2}^2 } .
\end{eqnarray}
The corresponding line element is expressed as:
\begin{eqnarray}
d\ell^2
  =\frac{ dv_{1}^2 + dv_{2}^{2}+(v_{1}dv_{2}-v_{2}dv_{1})^2 }{(v_{1}^{2}+v_{2}^{2}-1 )^2}
 = 4\,\frac{dzd\bar{z}}{(z\bar{z}-1)^2} .
  \label{eq:line-element4}
\end{eqnarray}

The second expression is the standard line element in the Klein disc model, and the third is essentially identical to the line element (\ref{eq:line-element3}), albeit with a factor of 4. 
This factor reflects the scale difference between the Poincar\'{e} and Klein models.
This leads to another solution that is isometrically distinct from the previous one. 
By substituting $(z_{1}, z_{2})$ with $(\xi_{1}(x), \xi_{2}(x))$, we obtain:
\begin{eqnarray}
(\xi(x), \eta(x)) = \left(\,\frac{ 2\xi_{1}(x) }{ 1+\xi_{1}(x)^2+\xi_{2}(x)^2 } \, , \,
                            \frac{ 2\xi_{2}(x) }{ 1+\xi_{1}(x)^2+\xi_{2}(x)^2 } \,\right).
  \label{eq:map(2)}
\end{eqnarray}
From the relation~(\ref{eq:Relation2}), the corresponding potentials are given as follows: 
\begin{eqnarray}
E
&=& 
 \frac{(\xi_{1}+1)^2+\xi_{2}^2}{(\xi_{1}-1)^2+\xi_{2}^2},  \\
F
&=& 
 \frac{2\xi_{2}}{(\xi_{1}-1)^2+\xi_{2}^2}.
\end{eqnarray}
Then, using the definition~(\ref{eq:Ernst pot}), we can finally obtain the following expressions:
\begin{eqnarray}
 e^{2\psi}
  &=& \left(\frac{1 - \xi_{1}^2 - \xi_{2}^2}{(\xi_{1}-1)^2+\xi_{2}^2}\right)^2,  \nonumber    \\
 \Phi,
  &=& 0,   \nonumber  \\
 A_{z}
 &=& \frac{2\xi_{2}}{(\xi_{1}-1)^2+\xi_{2}^2},    \nonumber   \\
\chi
&=& 0.
  \label{eq:setofsolution-2}
\end{eqnarray}

\subsubsection{\bf Seed harmonic maps treated here}

We examine two distinct types of maps that correspond to solutions of the vacuum Ernst equation (\ref{eq:VacuumErnst}). 
The first seed map, denoted as~(i), is derived from a specific set of geodesics in the target space and has been previously utilized in the previous works \cite{MishimaTomizawa2017,MT2022}. The second seed map, (ii), is based on the solitonic vacuum solution provided by Economou and Tsoubelis~\cite{Economou}.

\medskip
The expression of the seed map~(i) is given, as follows
\begin{eqnarray}
\xi_{v}(x) = \frac{1-e^{-2\tau(x)}+iA}{1+e^{-2\tau(x)}-iA} 
= \frac{1-A^2-e^{-4\tau(x)}}{(1+e^{-2 \tau(x)})^2+A^2}+i \frac{2A}{(1+e^{-2 \tau(x)})^2+A^2},
  \label{eq:MT17-1}
\end{eqnarray}
where $A$ is a real constant, and $\tau(x)$ is a real, cylindrically symmetric wave function that satisfies the linear wave equation $\nabla^2\tau = 0$ on ${\rm M^{(1,2)}}$.
On the other hand, the expression of the second seed map  (ii) is provided as follows
\begin{eqnarray}
\xi_{v}(x) = \frac{1-i l}{px - iqy}=\frac{px+l qy}{p^2x^2+q^2y^2}
                                          -i \frac{l p x-qy}{p^2x^2+q^2y^2},
  \label{eq:ET-1}
\end{eqnarray}
where $(x, y)$ represent pseudo-spheroidal coordinates. The parameters ${l, p, q}$ in this expression are subject to the constraint $q^2 - p^2 - l^2 = 1$. 
The relations between the coordinates $(x, y)$ and $(t, \rho)$ are as follows:
\begin{eqnarray}
 &&t= xy,\quad  \rho =\sqrt{(x^2+1)(y^2-1)},    \nonumber    \\
&&x= \frac{\sqrt{2}\,t}{\sqrt{1+r^2-t^2+\sqrt{4t^2+(1+r^2-t^2)^2}}},\quad 
y= \frac{\sqrt{1+r^2-t^2+\sqrt{4t^2+(1+r^2-t^2)^2}}}{\sqrt{2}}.
  \label{eq:ET-2}
\end{eqnarray}

\subsection{The expressions of the solutions}

In this section, we outline the solutions derived using the methodologies associated with cases (a) and (b), utilizing the seed maps (i) and (ii) discussed in the previous section. These solutions exemplify the application of the discussed harmonic mapping techniques to generate specific, physically relevant solutions of the vacuum Ernst equation, demonstrating the versatility and effectiveness of these methods in gravitational and electromagnetic field studies.

\subsubsection{\bf Solutions for case~(a)}

Applying the method associated with case~(a) to seed map~(i), we achieve expressions as detailed below, some of which have already been utilized in~\cite{MT2022}.
By substituting Eq.~(\ref{eq:MT17-1}) into Eq.~(\ref{eq:setofsolution-1}), we derive the subsequent expressions for the solution:

\begin{eqnarray}
e^{2\psi}
 &=& \frac{1}
       {A^2 e^{2\tau} \cos^4 \theta + e^{-2\tau} (\cos^2 \theta +e^{2\tau} \sin^2 \theta)^2 }, \nonumber   \\
\Phi
 &=& -{A e^{2\tau}\cos 2\theta }\,e^{2\psi},   \ \ \ \ \ 
\chi
 =  \frac{1}{2} {A e^{2\tau}\sin 2\theta }\, e^{2\psi},   \nonumber \\
A_{z}
 &=& -\frac{1}{2} 
   { [ A^2 e^{2\tau}\cos^2 \theta + ( e^{-2\tau}-1)( \cos^2 \theta + e^{2\tau}\sin^2 \theta) ] \sin 2\theta}\,e^{2\psi}.
     \label{eq:sol-1} 
\end{eqnarray}
Furthermore, by substituting the expressions~(\ref{eq:sol-1}) into Eq.~(\ref{eq:potdef2}), we obtain the following relation for the gravitational twist potential $w$:
\begin{eqnarray}
(\del{t}w,\, \del{\rho}w )=
  -4A\cos^4\theta \,\rho \left(\,\del{\rho}\tau, \ \del{t}\tau \, \right),
  \label{eq:w and tau1}
\end{eqnarray}
where the integrability of this equation is ensured if the function $\tau$ satisfies the linear wave equation in $M^{1,2}$. This condition underscores the relationship between the gravitational twist potential and the underlying wave function $\tau$, illustrating how the characteristics of $\tau$ influence the gravitational field's configuration.
Then,  the function $w$ is given as follows:
\begin{eqnarray}
w
 &=& -4 A\cos^4\theta \int^{(t,\rho)} \rho (\del{\rho}\tau dt + \del{t}\tau d\rho)  \nonumber  \\
 &=& -4 A\cos^4\theta \int^{(t,\rho)}_{(t,0)} \rho (\del{\rho}\tau dt + \del{t}\tau d\rho),
  \label{eq:w and tau2}
\end{eqnarray}
where an integration constant has been chosen such that $w(t, 0) = 0$. This choice is valid if the seed function $\tau$ is regular across the spacetime, ensuring that the integration yields a well-defined gravitational twist potential $w$ consistent with the boundary conditions of the system.

\medskip
Once the twist function $w$ is determined, the gauge potential $A_{\phi}$ can be derived as follows:
Starting by transforming Eq.~(\ref{eq:potdef1}) into the equation for $(\partial_{t}(A_{\phi} + w A_{z}), \partial_{\rho}(A_{\phi} + w A_{z}))$, and utilizing Eqs.~(\ref{eq:sol-1}) and (\ref{eq:w and tau1}), we obtain the following expression:
\begin{eqnarray}
(\del{t}(A_{\phi}+w A_{z}),\, \del{\rho}(A_{\phi}+w A_{z}) )
&=& ( \rho e^{-2\psi}\del{\rho}\chi + \del{t}w A_{z},\ \rho e^{-2\psi}\del{t}\chi + \del{\rho}w A_{z} )
  \nonumber  \\
&=& 
  -\tan\theta\, \left(\, \del{t}w, \ \del{\rho}w \, \right).
  \label{eq:Aphi1}
\end{eqnarray}
Thus, the gauge potential $A_{\phi}$ can be obtained, up to an arbitrary constant, as follows:
\begin{eqnarray}
A_{\phi} &=& -w\,A_{z} - \tan\theta \,w.
  \label{eq:Aphi}
\end{eqnarray}

\medskip
Finally, to determine the quantity $\gamma$, we substitute the expressions from (\ref{eq:sol-1}) into Eqs.~(\ref{eq:CenergyFlux-1}) and (\ref{eq:CenergyFlux-2}). 
As a result, ${\cal E}$ and ${\cal F}$ take the following simplified forms:
\begin{eqnarray} 
{\cal E}=\rho\left[ (\del{t}\tau)^2+  (\del{\rho}\tau)^2 \right], \ \ \  {\cal F}= 2\rho\del{t}\tau \del{\rho}\tau.
  \label{eq:C-energy density flux}
\end{eqnarray}
From this, the metric function $\gamma$ can be represented as a line integral:
\begin{eqnarray}
\gamma &=& 
\int^{(t,\rho)}_{(0,0)} \left\{2\rho\del{t}\tau \del{\rho}\tau dt + \rho\left[ (\del{t}\tau)^2+  (\del{\rho}\tau)^2 \right] d\rho\right\}+const.,  \nonumber \\
&=& 
\int^{(t,\rho)}_{(t,0)} \rho\left[ (\del{t}\tau)^2+  (\del{\rho}\tau)^2 \right] d\rho. 
  \label{eq:gammadesn1}
\end{eqnarray}
The second equality for $\gamma$ is affirmed by the assumption that $\gamma(t,0) = 0$, as indicated after Eq.~(\ref{eq:Cenergy-1}). It is significant to note that the expression for $\gamma$ is independent of the parameters $A$ and $\theta$. Therefore, $\gamma$ maintains the same form as in the case when $A = \theta = 0$, aligning with the characteristics of the original Einstein-Rosen wave. 
This consistency corresponds to a fundamental property of $\gamma$, namely isometric invariance in the target space.

\medskip
Next, applying the same method to seed map (ii), we derive the following expressions by substituting Eq.(\ref{eq:ET-1}) into Eq.(\ref{eq:setofsolution-1}):
\begin{eqnarray}
e^{2\psi}
 &=& \frac{p^2(x^2+1)+q^2(y^2-1)}
       {p^2x^2+q^2y^2-2(px+l qy)\cos2\theta+(l^2+1)\cos^{2}2\theta}, \nonumber   \\
\Phi
 &=& -\frac{2(l px-qy)\cos2\theta}
       {p^2x^2+q^2y^2-2(px+l qy)\cos2\theta+(l^2+1)\cos^{2}2\theta}, \nonumber   \\
A_{z}
 &=& \frac{[px+l qy -(l^2+1)\cos2\theta]\sin 2\theta}
       {p^2x^2+q^2y^2-2(px+l qy)\cos2\theta+(l^2+1)\cos^{2}2\theta}, \nonumber   \\
\chi
 &=& \frac{(l px-qy)\sin2\theta}
       {p^2x^2+q^2y^2-2(px+l qy)\cos2\theta+(l^2+1)\cos^{2}2\theta}.
     \label{eq:ETsol-1} 
\end{eqnarray}
Then, utilizing Eqs.~(\ref{eq:potdef1}) and (\ref{eq:potdef2}), we can determine the remaining quantities as follows:
\begin{eqnarray}
w
&=&\frac{y-1}{p[p^2(x^2+1)+q^2(y^2-1)]}  
   \{ q(l^2+1)(y+1)     \nonumber \\
     &&\ \ +2[p^2 l(x^2+1)-q^2 l(y+1) - qpx(y+1) ]\cos2\theta + q(l^2+1)(y+1)\cos^{2}2\theta \, \},
   \nonumber \\
A_{\phi}
&=&\frac{(y-1)\sin2\theta}
             { p[p^2x^2+q^2y^2-2(px+l qy)\cos2\theta+(\ell^2+1)\cos^{2}2\theta] }
   [ -p^{2} lx^{2} +q^{2}ly     \nonumber \\
     && \ \ \ +qpx(y+1) -q (l^2+1)(y+1)\cos2\theta + (l^3+1)\cos^{2}2\theta\,].
     \label{eq:ETsol-2} 
\end{eqnarray}
Due to the isometric invariance of the metric function $\gamma$ in the target space, the C-energy $\gamma$ is identical to that presented in the work \cite{Economou},
\begin{eqnarray} 
\gamma=\frac{1}{2}\ln \frac{p^{2}(x^2+1)+q^2(y^2-1)}{p^2(x^2+y^2)}.
  \label{eq:C-energy2}
\end{eqnarray}
Consequently, the density ${\cal E}$ and flux ${\cal F}$ corresponding to the C-energy are directly given as follows:
\begin{eqnarray} 
{\cal E}&=& \del{\rho}\gamma =\rho\frac{(l^2+1)(y^2-x^2+2x^2 y^2)}{(x^2+y^2)[(p^2(x^2+1)+q^2(y^2-1)]},  \nonumber  \\
{\cal F}&=& \del{t}\gamma =2\frac{(l^2+1)xy(x^2+1)(y^2-1)}{(x^2+y^2)[p^2(x^2+1)+q^2(y^2-1)]}.
  \label{eq:C-energy density flux2}
\end{eqnarray}

\subsubsection{\bf Solutions for case~(b)}
For the seed map (i), by utilizing Eq.~(\ref{eq:MT17-1}), the functions in (\ref{eq:setofsolution-2}) can be written as
\begin{eqnarray}
e^{2\psi }
  &=& \left(\frac{e^{2\tau } }{ A^{2}e^{4\tau }+1 }\right)^2,   \nonumber    \\
A_{z}
 &=& \frac{1}{A}- \frac{1}{A(A^{2}e^{4\tau}+1)},   \nonumber   \\
\Phi
  &=& 0, \ \ \ \chi = 0.
  \label{eq:setofsolution(b)-1}
\end{eqnarray}
By setting the parameter $A$ to $\cos\theta \sin\theta$ and replacing the seed function $\tau$ with $(\tau' -\ln(\cos^2\theta))/2$, which also satisfies the linear wave equation in ${\rm M^{(1,2)}}$, the above equations transform as follows:
\begin{eqnarray}
e^{2\psi }
  &=& \left(\frac{e^{\tau' } }{ e^{2\tau' }\sin^2\theta+\cos^2\theta }\right)^2,  
\nonumber    \\
A_{z}
 &=& \tan\theta + \left[\cot\theta- 
  \frac{1}{\cos\theta\sin\theta}\frac{1}{(\tan^{2}\theta e^{2\tau'}+1)}\right],   \nonumber   \\
\Phi
  &=& 0, \ \ \ \chi = 0.
  \label{eq:setofsolution(b)-2}
\end{eqnarray}
The observation that the derived expressions from the seed map (i) are essentially the same as those in Eq.~(\ref{eq:sol-1}) with the parameter $A$ set to 0 indicates that no new solution emerges from this specific case.

\medskip
For the seed map (ii), derived from Eq.~(\ref{eq:ET-1}), we obtain the following expressions:
 \begin{eqnarray}
e^{2\psi }
  &=& \left[\frac{ p^{2}(x^{2}+1)+q^{2}(y^2-1) }
                { (px-1)^2+(qy-l)^2 }\right]^2,  \nonumber    \\
w&=&0,\ \ \  
\gamma
=  2\ln \frac{p^{2}(x^2+1)+q^2(y^2-1)}{p^2(x^2+y^2)},  \nonumber  \\
A_{z}
 &=& -2\frac{lpx-qy}{ (px-1)^2+(qy-l)^2 },   \nonumber   \\
\Phi
  &=& 0, \ \ \ \chi = 0.
  \label{eq:setofsolution-3}
\end{eqnarray}
This solution corresponds to the solution in (\ref{eq:ETsol-1}) when $\theta = 0$.
In this case, the metric function $\psi$ in (\ref{eq:setofsolution-3}) corresponds to twice the function $\psi$ in~(\ref{eq:ETsol-1}), and the gauge potential component $A_z$ is identical to the twist potential $\Phi$ in~(\ref{eq:ETsol-1}). 
This relationship may be seen as an instance of the Bonnor transformation~\cite{Bonnor:1966,Fischer:1979}, which is treated in~\cite{book exact solution, book colliding waves}.
A treatment similar to this case has been already presented in~\cite{Yazadjiev:2005wf}, but its aims are essentially different from ours.
Consequently, the C-energy $\gamma$ and related quantities for the solution~(\ref{eq:setofsolution-3}) can be straightforwardly determined. 
The C-energy for this solution is exactly four times the value on the right-hand side of Eq.~(\ref{eq:C-energy2}). 
Similarly, the associated quantities ${\cal E}$ and ${\cal F}$ are four times those on the right-hand side of (\ref{eq:C-energy density flux2}), respectively. 

%

\section{Analysis}\label{sec:4}

In this section, we will investigate the nonlinear properties of the Einstein-Maxwell system through the behavior of the exact solution. Our primary focus will be on the mode conversion phenomena between gravitational and electromagnetic waves as significant occurrences. In particular, we aim to clarify the process and extent of these conversions when waves incident near the cylinder axis intensify rapidly. Furthermore, we will examine the rate at which conversion phenomena occur after the incident waves on the axis travel to the far side.
After briefly summarizing the asymptotic behavior of spacetimes corresponding to the derived solutions, we will dedicate our discussion to the mode conversion phenomena exhibited by each solution.
%
\subsection{Asymptotics}\label{sec:4A}

Here, we present the asymptotic forms of the metric obtained from the previous discussions for three distinct scenarios: 
case~(a) with seed~(i), case~(a) with seed~(ii), and case~(b) with seed~(ii).
In the subsequent discussion, we define the function $F(x)$ as 
\begin{eqnarray}
F(x) = 2x + \sqrt{4x^2 + 1}. \label{eq:Fx}
\end{eqnarray}

\medskip

For case  (a)
 with seed~(i), a typical example we consider is the Weber-Wheeler-Bonnor (WWB) solution with a single peak. The explicit form of this WWB solution is given as follows~\cite{Ashtekar2:1997},
\begin{eqnarray}
\tau(t,\rho) 
&=&
\frac{c}{\sqrt{2}}
\left[
 \frac{\sqrt{4 a^2 t^2 + (a^2 + \rho^2 - t^2)^2}
      +a^2 + \rho^2 - t^2}
      {4 a^2 t^2 + (a^2 + \rho^2 - t^2)^2}
\right]^{1/2}.
    \label{eq:WWBeq} 
\end{eqnarray}
The metric derived in this context bears resemblance to the one previously presented in \cite{MishimaTomizawa2017}. Consequently, it can be anticipated that the spacetime structure is almost identical to that of the case discussed earlier.

\medskip
Near the axis $\rho = 0$ (where $t = \text{const.}$), the metric can be approximately expressed as:
\begin{eqnarray}
ds^2
&\simeq& 
e^{f} \left[(\cos^{2}\theta + e^{f}\sin^{2}\theta)^2+e^{2f}A^2\cos^{4}\theta \right] ^{-1}
dz^2 \nonumber  \\  
&&
+e^{-f} \left[(\cos^{2}\theta + e^{f}\sin^{2}\theta)^2+e^{2f}A^2\cos^{4}\theta \right] \left( -dt^2 +d\rho^2+ \rho^2 d\phi^2 \right),
\end{eqnarray}
where $f= 2ac/(t^2+a^2)$. 
From the given metric expression, it becomes evident that the deficit angle around the axis disappears if the period of the angle $\phi$ is set to $2\pi$. This characteristic indicates that the spacetime near the axis is regular and free from angular deficits, which would otherwise suggest the presence of a conical singularity or similar topological defect.

\medskip
At spacelike infinity, where $\rho \to \infty$ (with $t = \text{const.}$), the asymptotic form of the metric is given by
\begin{eqnarray}
ds^2 \simeq \frac{1}{1 + A^2 \cos^4 \theta}dz^2 + \rho^2(1 + A^2 \cos^4 \theta)d\phi^2 + (1 + A^2 \cos^4 \theta)e^{\frac{c^2}{2a^2}}(-dt^2 + d\rho^2).
\end{eqnarray}
This indicates that the spacetime approaches a locally flat spacetime but with a deficit angle expressed as:
\begin{eqnarray}
\Delta\phi&=&2\pi-\lim_{\rho \to \infty}\frac{\int^{2\pi}_0\sqrt{g_{\phi\phi}}d\phi}{\int^\rho_0\sqrt{g_{\rho\rho}}d\rho}
               =2\pi \left(1-e^{-\frac{c^2}{4a^2}} \right).
    \label{eq:deficit} 
\end{eqnarray}

\medskip
At timelike infinities, as $t \to \pm\infty$ (with $\rho = \text{const.}$), the metric asymptotically approaches the form:
\begin{eqnarray}
ds^2 \simeq \frac{1}{1 + A^2 \cos^4 \theta}dz^2 + (1 + A^2 \cos^4 \theta)(-dt^2 + d\rho^2 + \rho^2 d\phi^2).
\end{eqnarray}
This indicates that at sufficiently late (or early) times, the spacetime converges to Minkowski spacetime, devoid of any deficit angle. 

\medskip
At future null infinity, where $v \to \infty$ ($u = \text{const.}$), the metric behaves as 
\begin{eqnarray}
ds^2&\simeq&\frac{1}{1+A^2\cos^4\theta}
\left(dz-2Ac\frac{F(-u)^{1/2}\cos^{4}\theta\sqrt{v}}{\sqrt{a^2+4u^2}  }d\phi \right)^2\nonumber\\
       &+&\rho^2(1+A^2\cos^4\theta)d\phi^2+(1+A^2\cos^4\theta)e^{F_+}(-dt^2+d\rho^2),
\end{eqnarray}
where
\begin{eqnarray*}
F_{+}=\frac{c^2[3a^4+12a^2u^2+32u^4-4u(a^2+4u^2)^{3/2} ] }{16a^2(a^2+4u^2)^2}.
\end{eqnarray*}

\medskip
At past null infinity, where $u \to -\infty$ ($v = \text{const.}$), the metric behaves as follows:
\begin{eqnarray}
ds^2&\simeq&\frac{1}{1+A^2\cos^4\theta}
\left(dz+2Ac\frac{F(v)^{1/2}\cos^{4}\theta\sqrt{-u}}{\sqrt{a^2+4u^2}  }d\phi \right)^2   \nonumber \\
       &+&\rho^2(1+A^2\cos^4\theta)d\phi^2+(1+A^2\cos^4\theta)e^{F_{-} }(-dt^2+d\rho^2),
\end{eqnarray}
where
\begin{eqnarray*}
F_{-}=\frac{c^2[3a^4+12a^2v^2+32v^4)+4v(a^2+4v^2)^{3/2} ] }{16a^2(a^2+4v^2)^2}.
\end{eqnarray*}
It is significant to note that $F_{+}$ (for $F_{-}$) reaches $0$ as $u$ (or $v$) approaches $\infty$ ($-\infty$), corresponding to timelike infinity, and attains the value of $c^2/4a^2$ as $u$ (or $v$) approaches $-\infty$ ($\infty$), corresponding to spacelike infinity. These findings indicate that spacetime rapidly becomes locally flat as we approach non-null infinities. Conversely, near null infinities, there remains a residual disturbance from waves. This pattern of behavior appears to be general. Indeed, the case involving solitonic waves demonstrates similar characteristics.


\medskip
For the case~(a) with the seed~(ii), which utilizes a solitonic solution as the seed, the asymptotic forms of the metric are as follows:
When the parameter $\theta$ is set to 0, the spacetime structure aligns with the vacuum spacetime described in~\cite{Economou}, with their parameter $|cp|=1$, indicating the vanishing condition of the deficit angle on the axis.

\medskip
Near the axis, the spacetime exhibits the following metric form:
\begin{eqnarray}
ds^2
&\simeq& 
G^{-1}dz^2+G \left( -dt^2 +d\rho^2+ \rho^2 d\phi^2 \right),
\end{eqnarray}
where
\begin{eqnarray}
G(t)=\frac{(l-q)^2+(1-pt)^2}{p^2(1+t^2)}
-4\sin^2\theta\frac{(l^2+1)\cos^2\theta-lq-pt}{p^2(1+t^2)}.
\end{eqnarray}

\medskip
At spacelike infinity, the metric asymptotically approaches the form:
\begin{eqnarray}
ds^2
&\simeq& 
\left\{dz+ 2\frac{(1+l^2-lq)+[-lq+(1+l^2)\cos^2\theta]\sin^2\theta}{pq} d\phi \right\}^2+ \frac{q^2}{p^2}\left( -dt^2 
+d\rho^2 \right)+ \rho^2 d\phi^2.
\end{eqnarray}
By appropriately redefining $z$, it can be shown that the spacetime approaches a flat spacetime with a deficit angle of $2\pi(1 -|q/p|)$.
Then, the metric form simplifies to:
\begin{eqnarray}
ds^2 &\simeq& dz^2 + \left( -dt^2 + d\rho^2 + \rho^2 d\phi^2 \right).
\end{eqnarray}
This indicates the emergence of Minkowski spacetimes.

\medskip
At future null infinity $v\to \infty$, the metric takes the form:
\begin{eqnarray}
ds^2&\simeq&
\left(dz+\frac{2\sqrt{F(-u)}(-lp+qF(-u))\sqrt{v}}{p^2+q^2(1-4u F(-u)) }\cos(2\theta)d\phi \right)^2     \nonumber   \\
       &+&\rho^2 d\phi^2+\frac{p^2+q^2(1-4u F(-u)) }{2p^2(1-2uF(-u))}(-dt^2+d\rho^2).
    \label{eq:futurenull(a)(ii)} 
\end{eqnarray}

\medskip
At past null infinity $u\to -\infty$, the metric behaves as:
\begin{eqnarray}
ds^2&\simeq&
\left(dz-\frac{2\sqrt{F(v)}(lp+qF(v))\sqrt{-u}}{p^2+q^2(1+4v F(v)) }\cos(2\theta)d\phi \right)^2     \nonumber   \\
       &+&\rho^2 d\phi^2+\frac{p^2+q^2(1+4v F(v)) }{2p^2(1+2vF(v))}(-dt^2+d\rho^2).
    \label{eq:pastnull(a)(ii)} 
\end{eqnarray}
It can be verified from Eqs.~(\ref{eq:futurenull(a)(ii)}) and~(\ref{eq:pastnull(a)(ii)}) that the metric coefficients $g_{\rho\rho}$ at future(past) null infinitiy approaches $1$ as $u(v) \to \infty (-\infty)$, corresponding to timelike infinity, while those take the value $q^2/p^2$ as $u(v) \to -\infty (\infty)$, corresponding to spacelike infinity.

\medskip
In discussing the case~(b) with the seed~(ii), we observe that the asymptotic forms of the spacetime are simple.

\medskip
Near the axis, the line element can be approximated as follows:
\begin{eqnarray}
ds^2 &\simeq& G_{0}^{-2}dz^2 + G_{0}^{2} \left( -dt^2 + d\rho^2 + \rho^2 d\phi^2 \right),
\end{eqnarray}
where $G_{0}$ is defined as $G|_{\theta=0}$.

\medskip
At spacelike infinity, the line element simplifies to:
\begin{eqnarray}
ds^2 &\simeq& dz^2 + \frac{q^8}{p^8}\left( -dt^2 + d\rho^2 \right) + \rho^2 d\phi^2.
\end{eqnarray}
Therefore, the corresponding spacetime approaches a flat spacetime with a deficit angle of $2\pi(1 - q^4/p^4)$.

\medskip
At the timelike infinities, the metric takes the form:
\begin{eqnarray}
ds^2 &\simeq& dz^2 + \left( -dt^2 + d\rho^2 + \rho^2 d\phi^2 \right).
\end{eqnarray}
This indicates the emergence of Minkowski spacetimes, similar to the case~(a) with seed~(ii).

\medskip
At future null infinity, the metric behaves as follows:
\begin{eqnarray}
ds^2 &\simeq& dz^2 + \rho^2 d\phi^2 + \left[\frac{p^2 + q^2(1 - 4u F(-u))}{2p^2(1 - 2uF(-u))}\right]^4 (-dt^2 + d\rho^2).
    \label{eq:futurenull(b)(ii)} 
\end{eqnarray}

\medskip
At past null infinity, the metric behaves as
\begin{eqnarray}
ds^2&\simeq& dz^2 +\rho^2 d\phi^2+\left[\frac{p^2+q^2(1+4v F(v)) }{2p^2(1+2vF(v))}\right]^4
(-dt^2+d\rho^2).
    \label{eq:pastnull(b)(ii)} 
\end{eqnarray}
As in the case~(a) with seed~(ii), from Eqs.~(\ref{eq:futurenull(b)(ii)}) and~(\ref{eq:pastnull(b)(ii)}) 
the metric coefficients $g_{\rho\rho}$ at future (past) null infinity approaches $1$ as $u (v) \to \infty (-\infty)$, while those take the value $q^8/p^8$ as $u (v) \to -\infty (\infty)$.

\subsection{Conversion between gravitational and electromagnetic waves}

\begin{figure}[h]
  \begin{tabular}{ccc}
 \begin{minipage}[t]{0.30\hsize}
 \centering
\includegraphics[width=5cm]{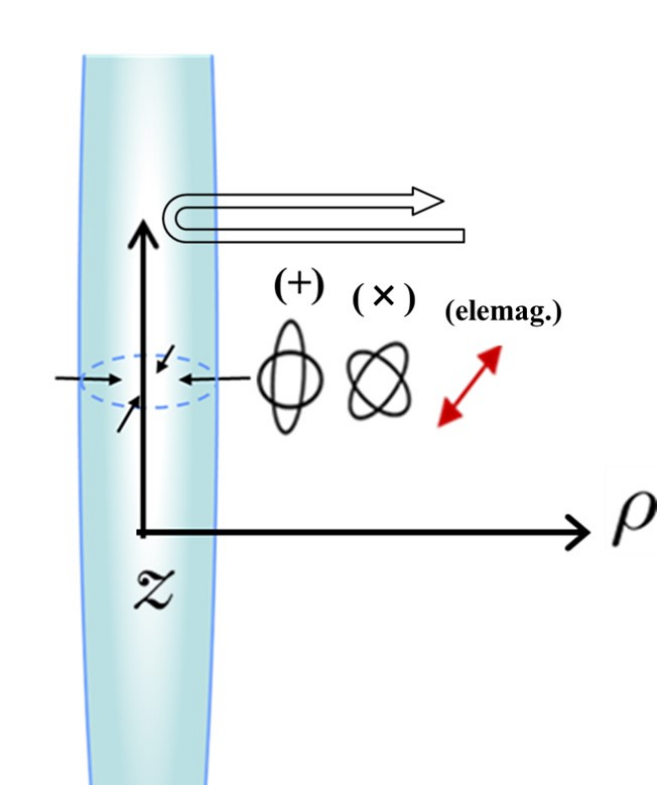}
 \end{minipage} &\ \ \ \ \ \ \ \ \ \ 
 \begin{minipage}[t]{0.30\hsize}
 \centering
\includegraphics[width=5cm]{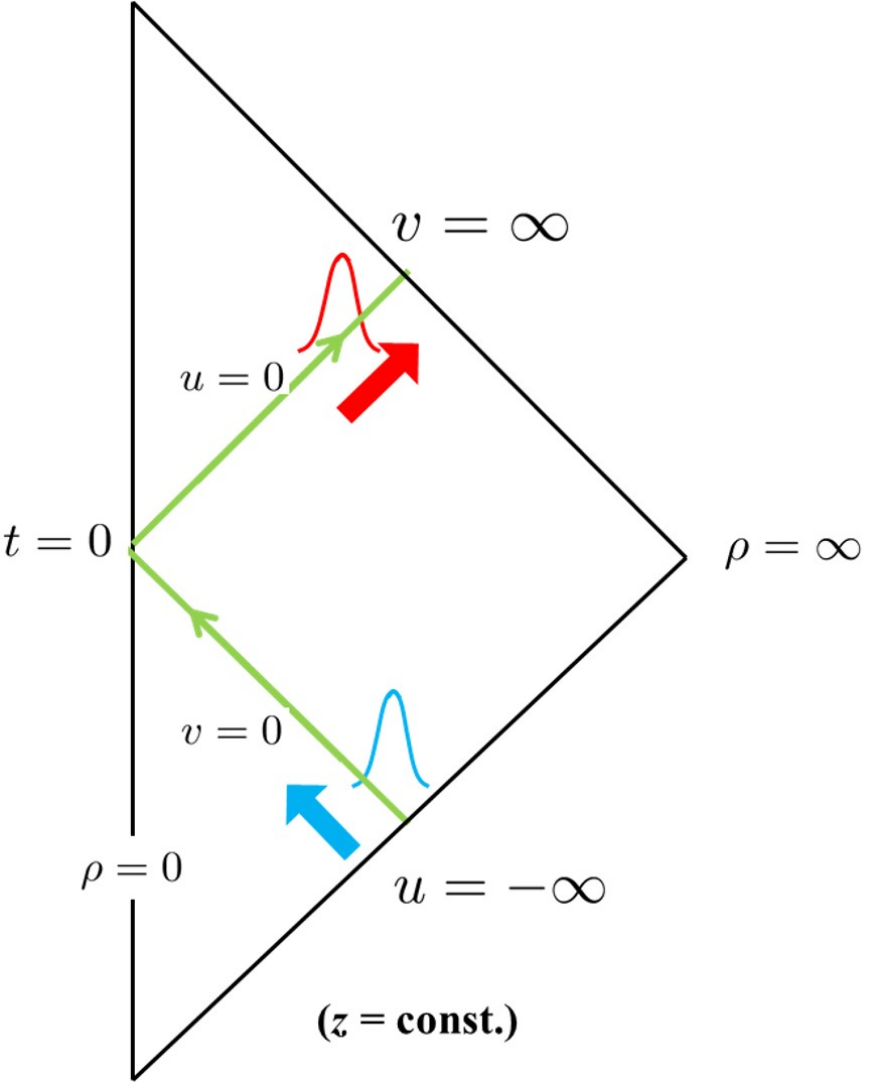}
 \end{minipage} &\ \ \ \ \ \ 
  \end{tabular}
\caption{
The left and right figures are schematic diagrams to make the setup clearer to understand. 
In the left figure, the cylindrical surface represents a wavefront.
The figure on the right is a kind of conformal diagram restricted to surfaces of $z={\rm const.}$.
 }
\label{fig:fig1}
\end{figure}

In the previous paper~\cite{MT2022}, we discussed the possibility of substantial mutual conversion between gravitational and electromagnetic modes in the spacetime corresponding to case~(a) seed~(i), as mentioned in the preceding section. Indeed, we observed that when a large cylindrically symmetric clump of gravitational waves, which includes a tiny portion of electromagnetic waves, explodes, the electromagnetic wave component is significantly amplified due to the vast amount of energy from the gravitational waves. As a continuation of this analysis, we use the exact solutions presented earlier to clarify the observed conversion phenomena. To introduce an experimental perspective (akin to a simulation), we consider scenarios resembling wave scattering experiments commonly studied in physics, as depicted in Fig.~\ref{fig:fig1}.

\medskip
In the first half of this section, following our previous work, we will examine case~(a) with seed~(i) in greater detail. We will explore a wider range of parameters than in the previous work, observing how the ratio of gravitational to electromagnetic modes contributing to the C-energy density changes over time as the waves incident near the axis scatter away. In the second half, we will address solitonic solutions for cylindrical waves within the Einstein-Maxwell system from the above perspective.

\subsubsection{\bf Case~(a) with seed~(i)}

\medskip
First, we introduce specific formulas for the occupancy ratio: each mode's contribution to the C-energy density ${\cal E}$, which is defined by Eq.~(\ref{eq:CenergyFlux-1}).
By using the expressions of physical quantities~(\ref{eq:sol-1}), the following formulas of the occupancy ratio are given:
\begin{eqnarray}
R_{\rm +}&:=&\frac{{\cal E}_{\rm +}}{{\cal E}}= 
  D^{-2} [\, (A^2e^{4\tau}-1)\cos^4\theta + e^{4\tau}\sin^4\theta \,]^2\,,  \nonumber  \\
R_{\rm \times}&:=&\frac{{\cal E}_{\rm +}}{{\cal E}}=
  4A^2 D^{-2}  e^{4\tau}\cos^8\theta\,,   \nonumber \\
R_{\phi}&:=& \frac{{\cal E}_{\phi}}{{\cal E}}= 
  4 A^2 D^{-3} e^{6\tau}( \cos^2\theta + e^{2\tau}\sin^2\theta  )^2\cos^4\theta \sin^2(2\theta)\,,
  \nonumber  \\
R_{z}&:=& \frac{{\cal E}_{z}}{{\cal E}}= 
  D^{-3} e^{2\tau}[\, A^2e^{4\tau} \cos^4\theta -( \cos^2\theta + e^{2\tau}\sin^2\theta )^2 \,]^2 \sin^2(2\theta)
  \,,
  \label{eq:occupancyEachMode}
\end{eqnarray}
where $D$ is 
$
{ A^2e^{4\tau}\cos^4\theta + ( \cos^2\theta + e^{2\tau}\sin^2\theta )^2 } 
$
and ${\cal E}$ is $\rho\left[ (\del{t}\tau)^2+ (\del{\rho}\tau)^2 \right]$ from Eq.(\ref{eq:C-energy density flux}).
The above quantity $R_{I}:={\cal E}_{I}/{\cal E}$ represents the occupancy for each mode $I$.
It should be noted that the C-energy density ${\cal E}$ itself has no dependence on individual modes, so that the mode contributions arise only from the occupancies.
From these formulas, the occupancies of gravitational and electromagnetic modes have been directly derived, as follows:
\begin{eqnarray}
R{\rm g} &:=& R_{+} + R_{\times}=
D^{-1}\left[A^2e^{4\tau}\cos^4\theta + ( \cos^2\theta - e^{2\tau}\sin^2\theta )^2 \right],
\nonumber \\
R_{\rm em} &:=& R_{\phi} + R_{z}=
D^{-1} e^{2\tau}\sin^2 (2\theta),
\label{eq:occupancyG-EM}
\end{eqnarray}
which have been used in the previous work~\cite{MT2022}.

\begin{figure}[h]
  \begin{tabular}{c}
\hspace{-10mm} \begin{minipage}[t]{1.0 \hsize}
 \centering
\includegraphics[width=15cm]{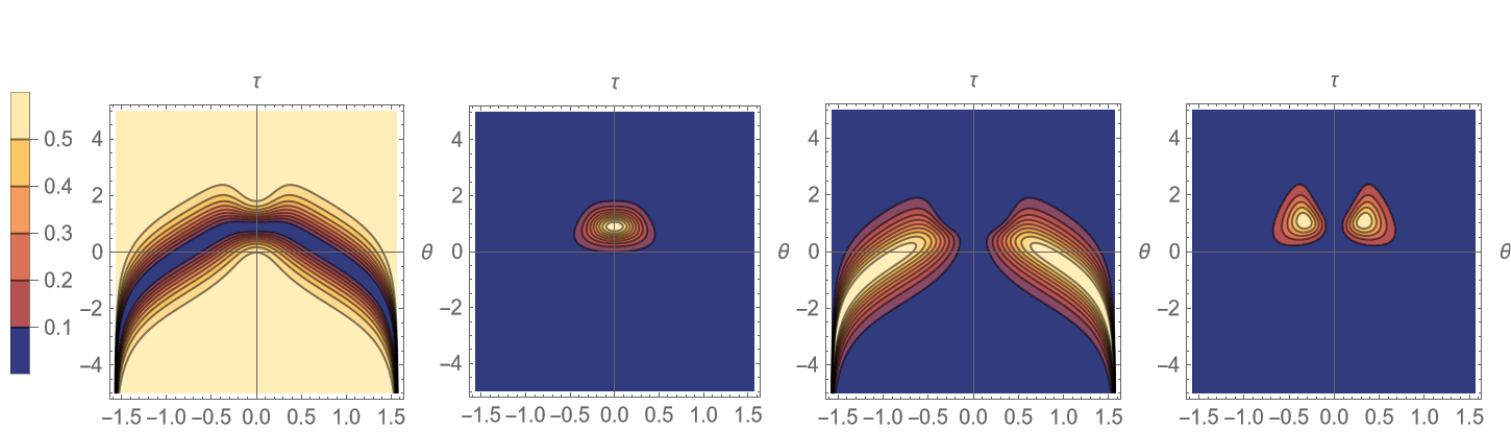}
 \end{minipage}
  \end{tabular}
\caption{
The case $A=1/6$: Four figures, from left to right, are graphs of contour plot of occupancy ratio of each mode: 
$+$, $\times$, $A_z$, $A_{\phi}$, respectively. 
}
\label{fig:fig2}
\end{figure}

\medskip
We now introduce a two-dimensional extension of the occupancy diagram used in our previous work. For instance, a set of diagrams for $A=1/6$ is depicted in Fig.~\ref{fig:fig2}. These contour plots are generated using Mathematica. The horizontal and vertical axes represent the parameter $\theta$ and the value of the seed function $\tau$, respectively. Each contour plot corresponds to a specific occupancy ratio $R_I$, where the subscript $I$ denotes one of the modes ${+, \times, A_z, A_{\phi}}$. By fixing the parameter $\theta$ and specifying the seed function $\tau$, we can determine the percentage of each mode's contribution at any point in spacetime from these diagrams. Furthermore, they allow us to qualitatively predict how mode conversion will occur. Indeed, the graph resulting from setting $\theta$ to a specific constant in this diagram is identical to the occupancy diagram treated in our previous work. However, these diagrams offer a broader perspective on how the conversion phenomenon varies as the parameter $\theta$ changes.

\begin{figure}[h]
  \begin{tabular}{c}
\hspace{0mm} \begin{minipage}[t]{1.0\hsize}
 \centering
\includegraphics[width=15cm]{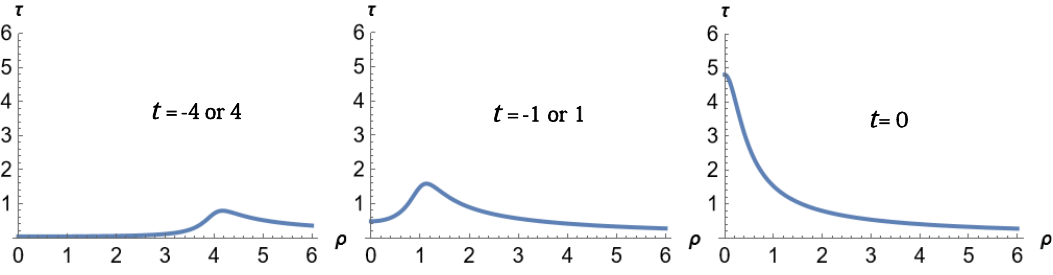}
 \end{minipage}
  \end{tabular}
\caption{
Snapshots of the WWB type seed function $\tau$: The parameters ($a$, $c$) are set to (1/3, 8/5). The function $\tau$ is an even function in time.
}
\label{fig:fig3}
\end{figure}

\medskip
To further our investigation, let us adopt the WWB impulse solution (referenced in Eq.~(\ref{eq:WWBeq})) as the seed function $\tau$. This same seed function has already been used in previous works. The width and height of the function $\tau$ are determined by the parameters $a$ and $c$. As an example, Fig.~\ref{fig:fig3} shows the case where $a=1/3$ and $c=8/5$. The graphs in Fig.~\ref{fig:fig3} demonstrate that the wave height increases as the wave approaches the cylindrical axis ($\rho=0$) and decreases with increasing distance.

\begin{figure}[h]
  \begin{tabular}{c}
\hspace{-5mm} \begin{minipage}[t]{1.0\hsize}
 \centering
\includegraphics[width=15cm]{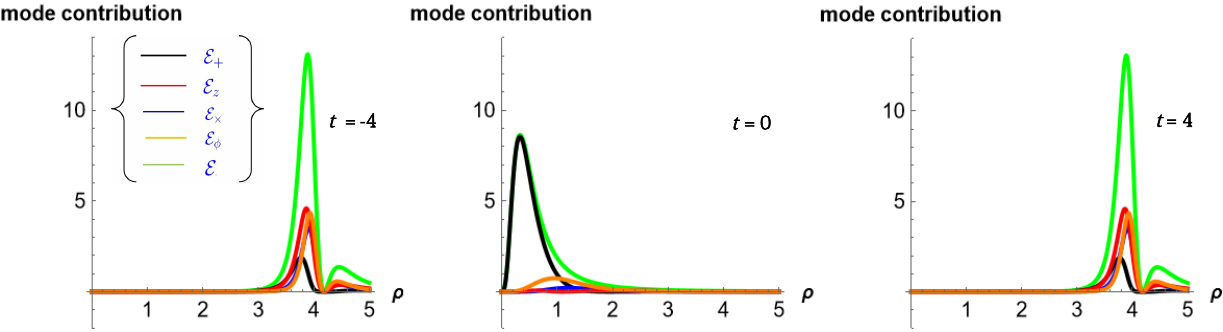}
 \end{minipage}
  \end{tabular}
\caption{
Snapshots of each mode contribution for the case of $( A, \theta, a, c )=(1/6, \pi/10, 1/3, 8/5)$: Each line of black, red, blue, yellow, and green 
corresponds to $+$, $A_{z}$, $\times$, $A_{\phi}$, and total C-energy density, respectively. 
At $t=\pm 4$ the blue line is hidden by the red and yellow lines.
}
\label{fig:fig4}
\end{figure}

\medskip
In general, it is reasonable to assume that a cylindrically symmetric wave packet, when launched toward the symmetric axis from a distance, initially has a small peak which grows as it approaches the axis and diminishes as it moves away. 
It can be known by examining the representation of Eq.(\ref{eq:C-energy density flux}) that the C-energy is generally concentrated near the peak of the wave packet.
 if the parameter $\theta$ is predefined, we can track how the mode conversion evolves over time by referencing the corresponding diagrams in Fig.~\ref{fig:fig2} (also see~\cite{MT2022}). For instance, if $\theta$ is set to $\pi/10$, corresponding to a choice of a vertical line on the diagram, the peak of the seed function $\tau$ begins near $\tau = 0$ (i.e., near past null infinity), and either ascends or descends along the vertical line (the direction depends on the sign of the $\tau$ peak). After reaching the maximum of $|\tau|$ (i.e., at the symmetric axis), it returns to the starting point. Concurrently, as $\tau$ progresses along the line, the contour plots in each diagram indicate the occupancy of the corresponding mode at every point the wave packet traverses. The snapshots in Fig.\ref{fig:fig4} display the temporal behaviors of mode contributions to C-energy density for the case: $(A, \theta, a, c) = (1/6, \pi/10, 1/3, 8/5)$. This figure shows that at $t=\pm 4$, ${\cal E}_{z}$, ${\cal E}_{\times}$, and ${\cal E}_{\phi}$ dominate over ${\cal E}{+}$. Conversely, at $t=0$, the $+$ mode temporarily becomes more dominant than the other modes. This observation can be inferred by carefully seeing the occupancy diagrams in Fig.~\ref{fig:fig2}.

\begin{figure}[h]
  \begin{tabular}{c}
\hspace{-10mm} \begin{minipage}[t]{1.0 \hsize}
 \centering
\includegraphics[width=15cm]{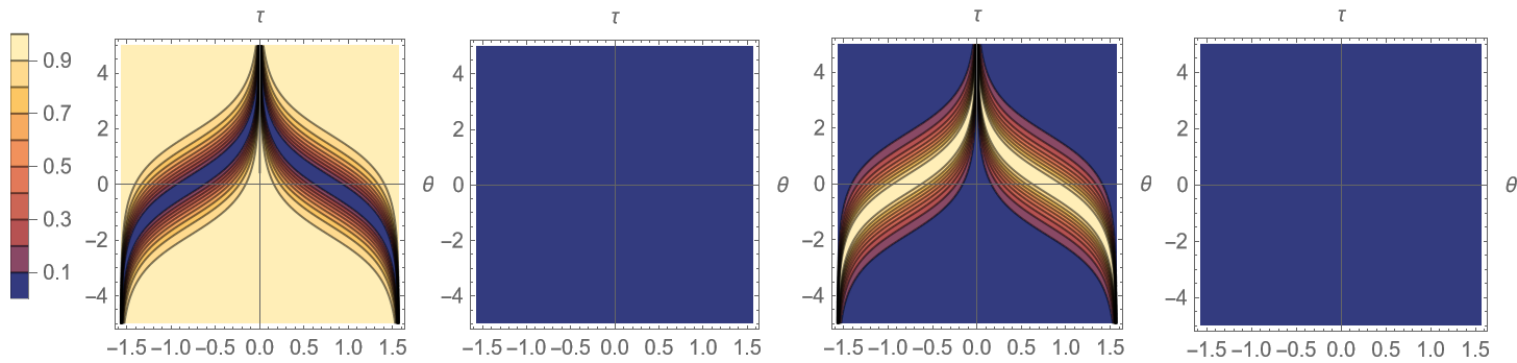}
 \end{minipage}
  \end{tabular}
\caption{
The case $A=0$: Four figures, from left to right, are graphs of contour plot of occupancy ratio of each mode: $+$, $\times$, 
$A_z$, $A_{\phi}$, respectively. 
}
\label{fig:fig5}
\end{figure}

\medskip
Next, let us consider the simple but intriguing case where $A=0$, in which neither the gravitational $\times$ nor the electromagnetic $A_{\phi}$ modes are excited. This fact is confirmed directly by the expressions in Eq.~(\ref{eq:occupancyEachMode}). 
The occupancy diagram illustrating this is shown in Fig.~\ref{fig:fig5}. Indeed, the second and fourth figures within this diagram display no signs of excitation for the $\times$ mode and $A_{\phi}$ mode, respectively.

\begin{figure}[h]
  \begin{tabular}{c}
\hspace{-10mm} \begin{minipage}[t]{1.0 \hsize}
 \centering
\includegraphics[width=15cm]{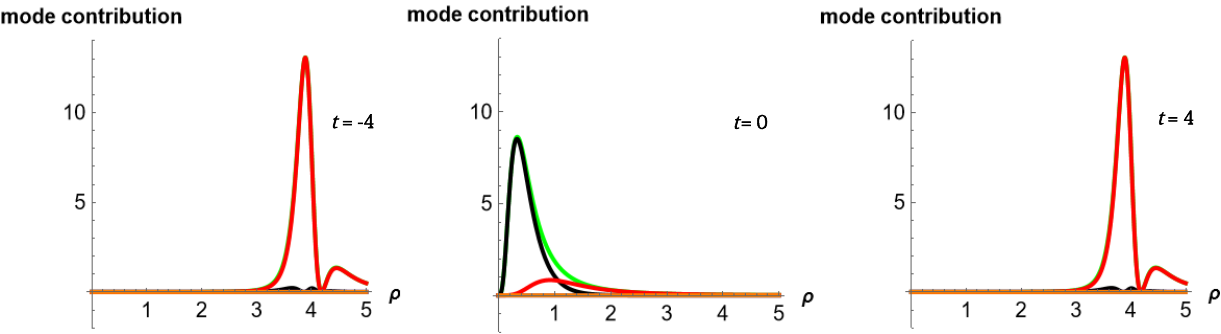}
 \end{minipage}
  \end{tabular}
\caption{
Snapshots of each mode contribution for the case of $( A, \theta, a, c )=(0, \pi/6, 1/3, 8/5)$: Each line of black, red, and green 
corresponds to $+$, $A_{z}$, and total C-energy density, respectively. 
At $t=\pm 4$ the green line is almost completely hidden by the red line.
}
\label{fig:fig6}
\end{figure}

\medskip
By examining the diagrams, we can identify some parameters $\theta$ that correspond to waves with characteristic behavior. For example, the behavior along the line at $\theta=\pi/6$ suggests the following dynamics: initially, when the wave is positioned far from the axis (i.e., near past null infinity), it predominantly exhibits electromagnetic contributions. As the wave moves away from this position and approaches the axis, its strength increases. Upon nearing the axis, the wave primarily transitions to gravitational modes. The snapshots in Fig.~\ref{fig:fig6} display the temporal behavior of these corresponding waves. It is evident from this figure that at $t=\pm 4$, the wave is predominantly in the electromagnetic $A_{z}$ mode. At $t=0$, as the wave reaches the axis, there is significant amplification of the gravitational $+$ mode.

\begin{figure}[h]
  \begin{tabular}{c}
\hspace{-10mm} \begin{minipage}[t]{1.0 \hsize}
 \centering
\includegraphics[width=15cm]{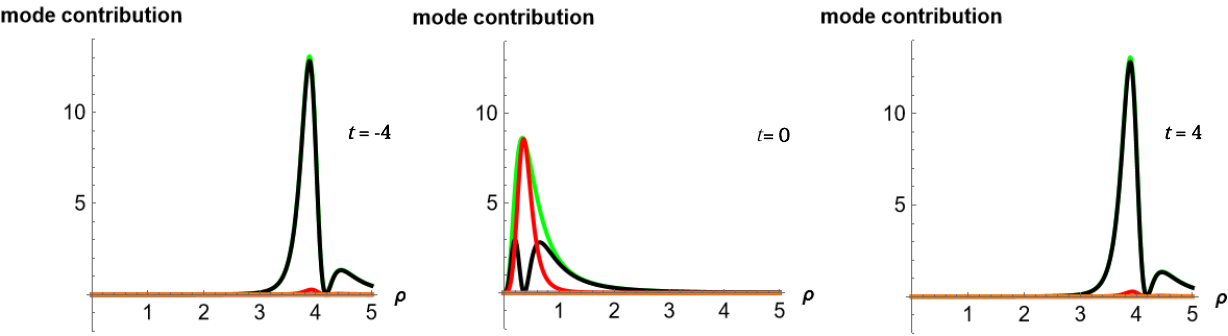}
 \end{minipage}
  \end{tabular}
\caption{
Snapshots of each mode contribution for the case of $( A, \theta, a, c )=(0, 11\pi/24, 1/3, 8/5)$: Each line of black, red, and green 
corresponds to $+$, $A_{z}$, and total C-energy density, respectively. 
At $t=\pm 4$ the green line is almost completely hidden by the black line. respectively. 
}
\label{fig:fig7}
\end{figure}

\medskip
As another example, consider the case $\theta = 11\pi/24$ ($\approx 1.44$). In this case, the trajectory along which the wave propagates is positioned near the left edge of the diagram. From this placement, it is easily anticipated that if the sign of the seed $\tau$ is negative, the initially small electromagnetic wave accompanying the large gravitational wave will be greatly amplified near the axis. Indeed, the snapshots in Fig.~\ref{fig:fig7} clearly illustrate the occurrence of this conversion phenomenon as described.

\medskip
Before proceeding, let us discuss whether the wave retains its altered occupancy after leaving the vicinity of the axis (i.e., the strongly nonlinear region) and returning to the original distant side from the axis (i.e., near future null infinity, as shown in Fig.\ref{fig:fig1}). In conclusion, if the seed $\tau$ is a decreasing function at infinity, the occupancy rates evaluated at both past and future null infinities always coincide, as indicated by the formulas in Eq.~(\ref{eq:occupancyEachMode}). This means that in this category of solutions, the reflected wave inevitably returns with the same content as the initial wave, even if the seed $\tau$ is not an even function with respect to $t$. This observation naturally leads to the following question: to what extent do nontrivial conversions, particularly between gravitational and electromagnetic modes, occur when the incident and reflected waves are at sufficiently far distances? 
This issue~\cite{Harada} prompts further investigation. Let us now continue by analyzing the remaining two cases to seek answers to this question.

\subsubsection{\bf Case~(a) with seed~(ii)}

To delve into the question posed, we investigate the solution for case~(a) with seed~(ii). We start by providing the expressions for the mode contributions to the C-energy density at past or future null infinity, corresponding to 
 $u =-\infty$ or $v=\infty$, respectively.
The expressions are given below:
\begin{eqnarray}
{}^{(-)}{\cal E}_{+}(v|p,q,l) &=& \cos^2(2 \theta)\, {\cal K}_{+}(v|p,q,l), \ \ \ \ 
{}^{(-)}{\cal E}_{\times}(v|p,q,l) = \cos^2(2 \theta)\, {\cal K}_{\times}(v|p,q,l),  \nonumber   \\
{}^{(-)}{\cal E}_{z}(v|p,q,l) &=& \sin^{2}(2 \theta)\, {\cal K}_{+}(v|p,q,l),    \ \ \ \ 
{}^{(-)}{\cal E}_{\phi}(v|p,q,l) = \sin^{2}(2 \theta)\,  {\cal K}_{\times}(v|p,q,l),  \nonumber   \\
{}^{(+)}{\cal E}_{+}(u|p,q,l) &=&  \cos^2(2 \theta)\, {\cal K}_{+}(-u|p,q,-l) ,\ \ \ \ 
{}^{(+)}{\cal E}_{\times}(u|p,q,l) = \cos^2(2 \theta)\, {\cal K}_{\times}(-u|p,q,-l),  \nonumber   \\
{}^{(+)}{\cal E}_{z}(u|p,q,l) &=& \sin^{2}(2 \theta)\, {\cal K}_{+}(-u|p,q,-l),  \ \ \ \ 
{}^{(+)}{\cal E}_{\phi}(u|p,q,l) = \sin^{2}(2 \theta)\,  {\cal K}_{\times}(-u|p,q,-l),
           \label{eq:Infoccupancyeachmode(a)(ii)}
\end{eqnarray}
where superscripts $(-)$ and $(+)$ are assigned to past and future null infinities, respectively, and the functions ${\cal K}_{+}$ and ${\cal K}_{\times}$ are defined as follows.
\begin{eqnarray}
{\cal K}_{+}(x|p,q,l)
&:=&
\frac{F(x) \left\{q F(x) \left[-12 p q x-3 l p^2+l q^2 \left(16 x^2+1\right)\right]+4 l q^3 x+p^3-3 p q^2 \right\}^2}
       {2 \left(4 x^2+1\right) \left(4 q^2 x F(x)+p^2+q^2 \right)^4},        \nonumber \\
{\cal K}_{\times}(x|p,q,l)
&:=&
\frac{F(x) \left\{ q F(x) \left[12 l p q x-3 p^2+q^2 \left(16 x^2+1 \right)\right]+4 q^3 x -l p^3 +3 l p q^2 \right\}^2}
       {2 \left(4 x^2+1\right) \left(4 q^2 x F(x)+p^2+q^2 \right)^4}, 
           \label{eq:Infdfunctiondef(a)(ii)}
\end{eqnarray}
where $F(x)$ has been already given in Eq.~(\ref{eq:Fx}) as a positive definite function. 
It is noteworthy that the functions ${\cal K}_{+}$ and ${\cal K}_{\times}$ satisfy the following useful formulas:
\begin{eqnarray}
{\cal K}_{I}(x|p,q,l)
= {\cal K}_{I}(x|-p,q,-l)= {\cal K}_{I}(x|p,-q,-l)= {\cal K}_{I}(x|-p,-q,l).
           \label{eq:FormulasforK}
\end{eqnarray}

\medskip
The gravitational and electromagnetic contributions to the C-energy density ${\cal E}$ are given as follows:
\begin{eqnarray}
{}^{(-)}{\cal E}_{\rm g}(v) &:=& {}^{(-)}{\cal E}_{+}(v) + {}^{(-)}{\cal E}_{\times}(v)
 = \frac{\left(l^2+1\right) \cos ^2(2 \theta) F(v)}{2 \left(4 v^2+1\right) \left(4 q^2 v F(v)+p^2+q^2\right)},  \nonumber \\
{}^{(-)}{\cal E}_{\rm em}(v) &:=& {}^{(-)}{\cal E}_{z}(v) + {}^{(-)}{\cal E}_{\phi}(v)
 = \frac{\left(l^2+1\right) \sin^2 (2 \theta) F(v)}{2 \left(4 v^2+1\right) \left(4 q^2 v F(v)+p^2+q^2\right)}, \nonumber  \\
{}^{(+)}{\cal E}_{\rm g}(u) &:=& {}^{(+)}{\cal E}_{+}(u) + {}^{(+)}{\cal E}_{\times}(u) = {}^{(-)}{\cal E}_{\rm g}(-u),  \nonumber  \\
{}^{(+)}{\cal E}_{\rm em}(u) &:=& {}^{(+)}{\cal E}_{z}(u)+ {}^{(+)}{\cal E}_{\phi}(u)= {}^{(-)}{\cal E}_{\rm em}(-u).
           \label{eq:InfoccupancyG-EM(a)(ii)}
\end{eqnarray}
More general expressions of contributions corresponding to gravitational and electromagnetic parts, defined at arbitrary points in spacetime, are also given as follows:
\begin{eqnarray}
{\cal E}_{\rm g}&=&
\frac{\cos ^2(2 \theta) \left(p^2 x^2+q^2 y^2\right)-2 \cos (2 \theta) (l q y+p x)+l^2+1}
{\left(l^2+1\right) \cos ^2(2 \theta)-2 \cos (2 \theta) (l q y+p x)+p^2 x^2+q^2 y^2}\,{\cal E},   \nonumber  \\
{\cal E}_{\rm em}&=&
\frac{\sin ^2(2 \theta) \left(p^2 x^2+q^2 y^2-l^2-1\right)}
{\left(l^2+1\right) \cos ^2(2 \theta)-2 \cos (2 s\theta) (l q y+p x)+p^2 x^2+q^2 y^2}\,{\cal E},
           \label{eq:GeneraloccupancyG-EM(a)(ii)}
\end{eqnarray}
where ${\cal E}$ has already been given in Eq.~(\ref{eq:C-energy density flux2}), and the coordinates $(x,y)$ can be transformed into $(\rho, t)$ by the formulas in (\ref{eq:ET-2}). Here, however, the detailed expressions of each mode contribution are omitted due to the complexity of the expressions. Whenever necessary, the explicit forms can be generated by substituting Eq.~(\ref{eq:ETsol-1}) into Eq.~(\ref{eq:amplitudes}) and then using the formula~(\ref{eq:CenergyFlux-1}).

\medskip
First, it should be noted from Eq.~(\ref{eq:InfoccupancyG-EM(a)(ii)}) that the gravitational contribution ${}^{(-)}{\cal E}_{\rm g}$ and the electromagnetic contribution ${}^{(-)}{\cal E}_{\rm em}$ are essentially the same as ${}^{(+)}{\cal E}_{\rm g}$ and ${}^{(+)}{\cal E}_{\rm em}$, respectively. This indicates that the occupancy of gravitational and electromagnetic modes in C-energy density is invariant when compared at past null infinity and future null infinity. It can be expected that even within the scenario of case~(a) with seed~(ii), the conversion between gravitational and electromagnetic waves does not occur. Additionally, it can be demonstrated from Eq.~(\ref{eq:Infoccupancyeachmode(a)(ii)}) that the forms of modes $+$($\times$) and $A_{z}$($A_{\phi}$) are exactly identical except for an overall constant factor.

\medskip
To further elaborate the analysis, let us examine the expressions for ${}^{(-)}{\cal E}_{I}$ and ${}^{(+)}{\cal E}_{I}$ in Eq.~(\ref{eq:Infoccupancyeachmode(a)(ii)}), where the subscript {\it I} denotes $+$, $\times$, $z$, and $\phi$. A key distinction from the previous case~(a) with seed~(i) is that for each mode ($I$), ${}^{(-)}{\cal E}_{I}$ differs from ${}^{(+)}{\cal E}_{I}$ except in the case $l=0$.

\begin{figure}[h]
  \begin{tabular}{cc}
\hspace{-10mm} \begin{minipage}[t]{0.8 \hsize}
 \centering
\includegraphics[width=12cm]{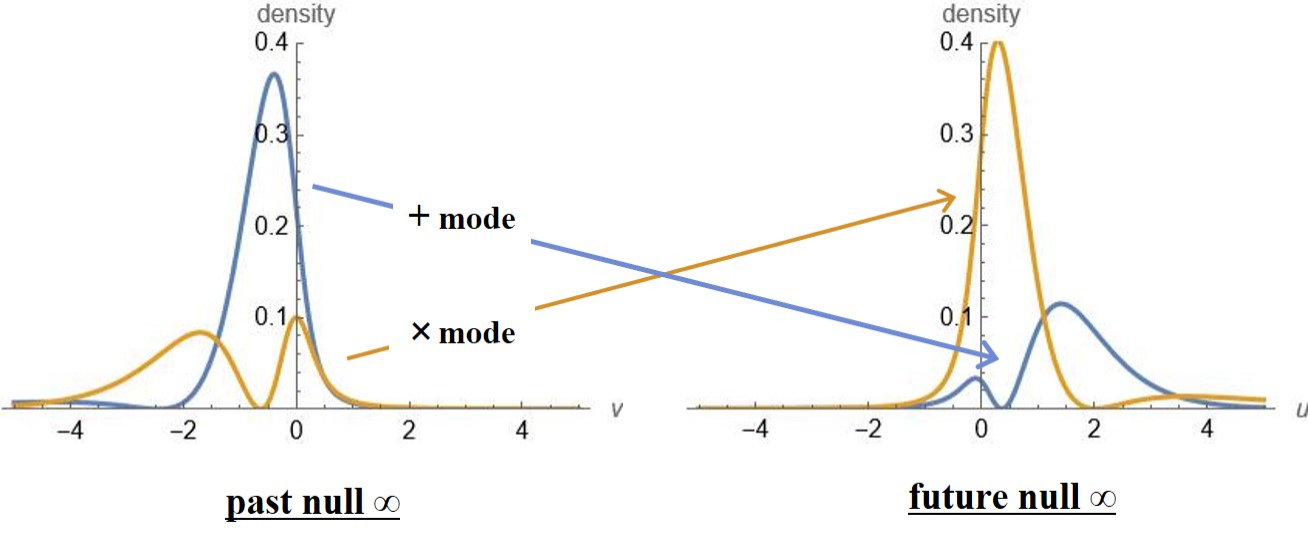}
 \end{minipage}
  \end{tabular}
\caption{
Conversion between gravitational $+$ (blue line) and $\times$ (yellow line) modes observed at the past and 
future null infinity: 
$(\theta, q, p, l) =(\pi/10, -9\sqrt{5}/16, \sqrt{5}/16, 3/4) $. Two additional arrows are superimposed for assistance.
}
\label{fig:fig8}
\end{figure}
\begin{figure}[h]
  \begin{tabular}{cc}
\hspace{-10mm} \begin{minipage}[t]{0.8 \hsize}
 \centering
\includegraphics[width=14cm]{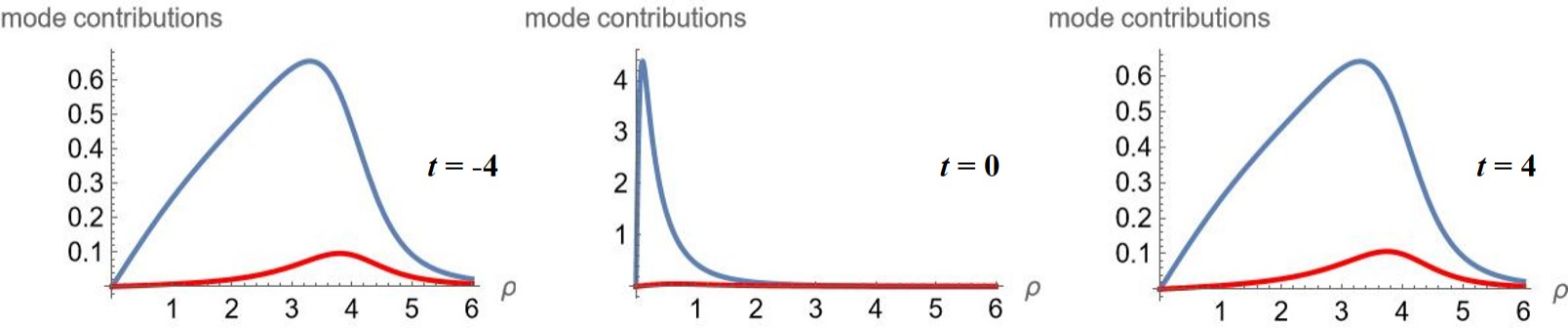}
 \end{minipage}
  \end{tabular}
\caption{
Snapshots at $t=-4, 0, 4$ of the gravitational(blue line) and electromagnetic(red line) contributions to C-energy density for  
$(\theta, q, p, l) =(\pi/10, -9\sqrt{5}/16, \sqrt{5}/16, 3/4) $
}
\label{fig:fig9}
\end{figure}
\begin{figure}[h!]
  \begin{tabular}{cc}
\hspace{-10mm} \begin{minipage}{0.8 \hsize}
 \centering
\includegraphics[width=11cm]{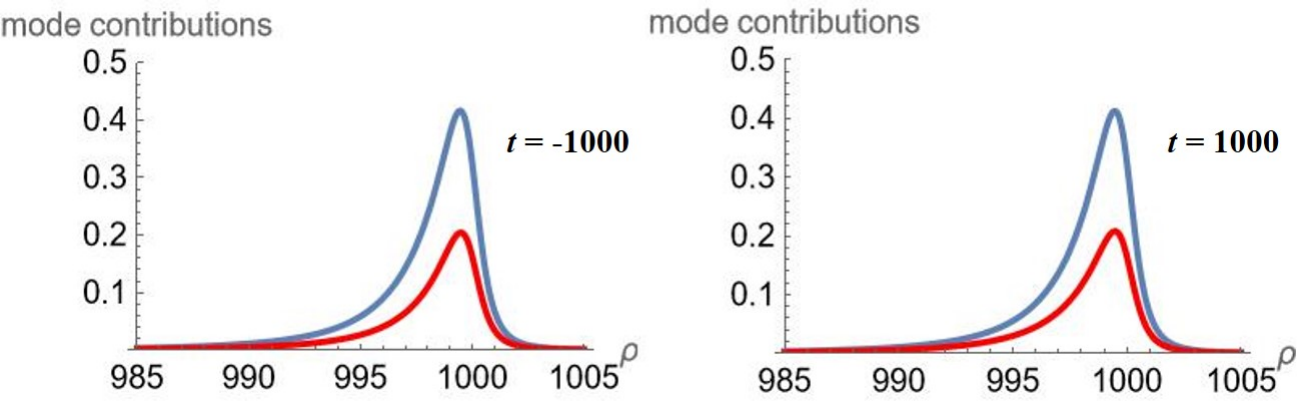}
 \end{minipage}
  \end{tabular}
\caption{
Snapshots at $t=-1000, 1000$ of the gravitational(blue line) and electromagnetic(red line) contributions to 
C-energy density for $(\theta, q, p, l) =(\pi/10, -9\sqrt{5}/16, \sqrt{5}/16, 3/4) $
}
\label{fig:fig10}
\end{figure}

\medskip
It can therefore be shown, using Eq.~(\ref{eq:Infoccupancyeachmode(a)(ii)}), that conversions within the gravitational or electromagnetic modes can occur. In other words, we can observe a type of ``Faraday rotation" within these solutions. In fact, Fig.~\ref{fig:fig8} presents an example of a conversion between gravitational $+$ and $\times$ modes as the wave moves from past null infinity toward the axis and is reflected back to future null infinity. As the additional arrows in Fig.~\ref{fig:fig8} indicate, a significant conversion within the gravitational component is clearly visible.
Conversely, according to Eq.~(\ref{eq:GeneraloccupancyG-EM(a)(ii)}), near the axis, the waves exhibit
a substantial depression of the electromagnetic modes
 due to the nonlinear effects in this region, as illustrated in Fig.~\ref{fig:fig9}. However, as expected and depicted in Fig.~\ref{fig:fig10}, the occupancies corresponding to the incident and reflected waves become closer as the observer's position moves away from the axis.

\subsubsection{\bf Case~(b) with seed~(ii)}

The third case (i.e., case~(b) with seed~(ii)) is not geometrically equivalent to the other cases treated above. Therefore, the wave represented by the solution corresponding to the third case can be expected to exhibit behaviors essentially different from those in other cases. In fact, comparing Eq.~(\ref{eq:ETsol-1}) with Eq.~(\ref{eq:setofsolution-3}), the expressions for $2\psi$ and $\Phi$ at $\theta=0$ in Eq.~(\ref{eq:ETsol-1}) are exactly equal to those for $\psi$ and $A_{z}$ in Eq.~(\ref{eq:setofsolution-3}). Hence, according to considerations similar to those for case~(b) with seed~(ii), the conversion between the gravitational and electromagnetic parts, evaluated at null infinities, can be expected to occur in a non-trivial manner.

\begin{figure}[h!]
  \begin{tabular}{cc}
\hspace{-10mm} \begin{minipage}{0.8 \hsize}
 \centering
\includegraphics[width=11cm]{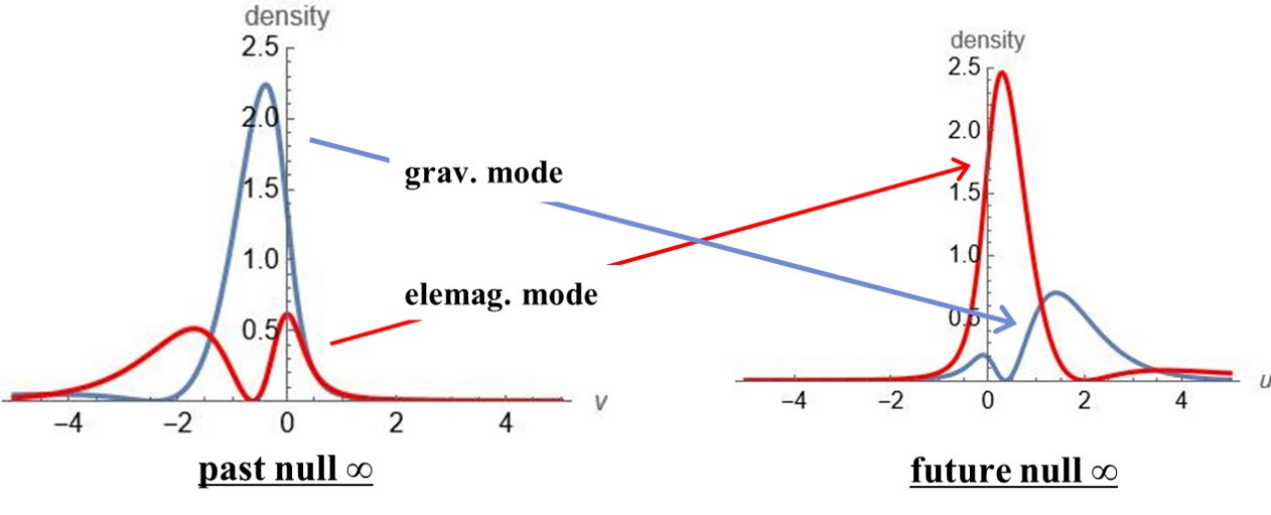}
 \end{minipage}
  \end{tabular}
\caption{
Conversion between gravitational (blue line) and electromagnetic (red line) modes observed at the past and 
future null infinity: $(q, p, l) =(-9\sqrt{5}/16, \sqrt{5}/16, 3/4) $. Two additional arrows are superimposed for assistance.
}
\label{fig:fig11}
\end{figure}

\begin{figure}[h!]
  \begin{tabular}{cc}
\hspace{-10mm} \begin{minipage}{0.8 \hsize}
 \centering
\includegraphics[width=14cm]{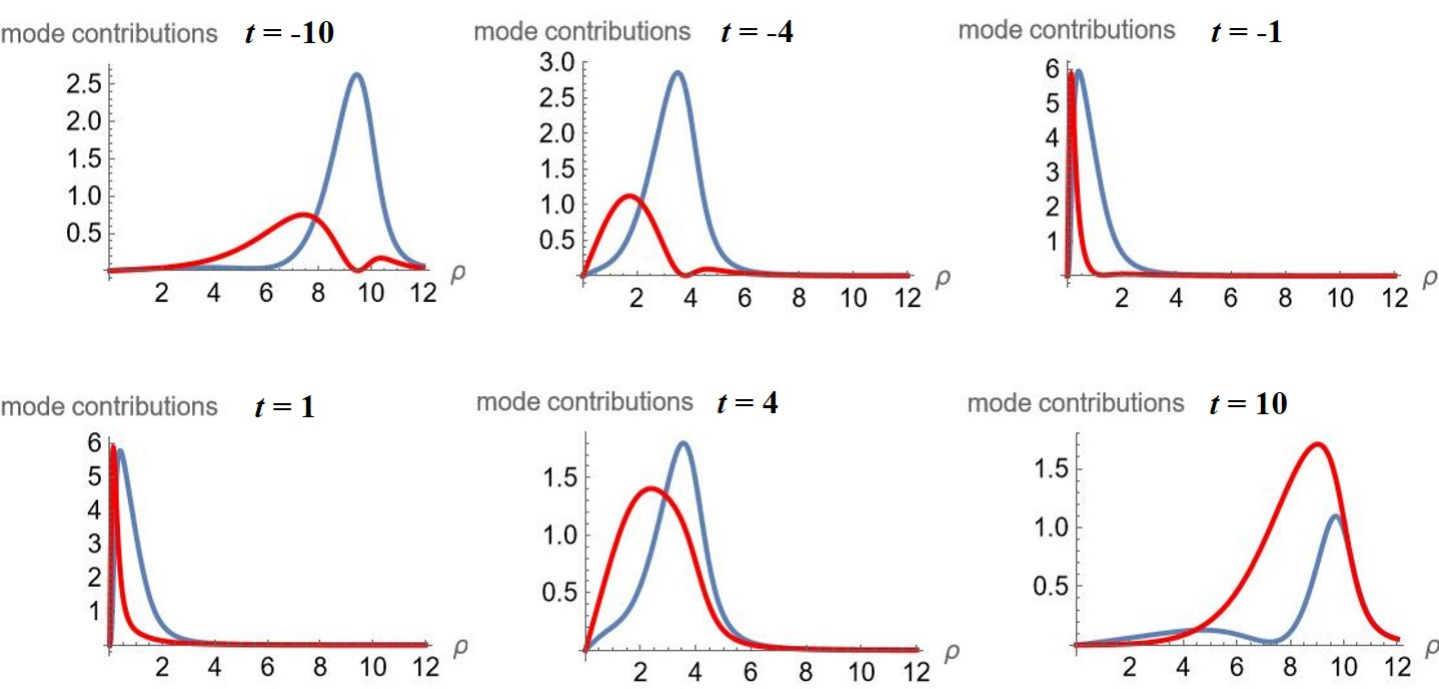}
 \end{minipage}
  \end{tabular}
\caption{
Snapshots at $t=\pm 10, \pm 4, \pm 1$ of the gravitational (blue line) and electromagnetic (red line) contributions to C-energy density for  
$(q, p, l) =(-9\sqrt{5}/16, \sqrt{5}/16, 3/4) $
}
\label{fig:fig12}
\end{figure}

\begin{figure}[h!]
  \begin{tabular}{cc}
\hspace{-10mm} \begin{minipage}{0.8 \hsize}
 \centering
\includegraphics[width=11cm]{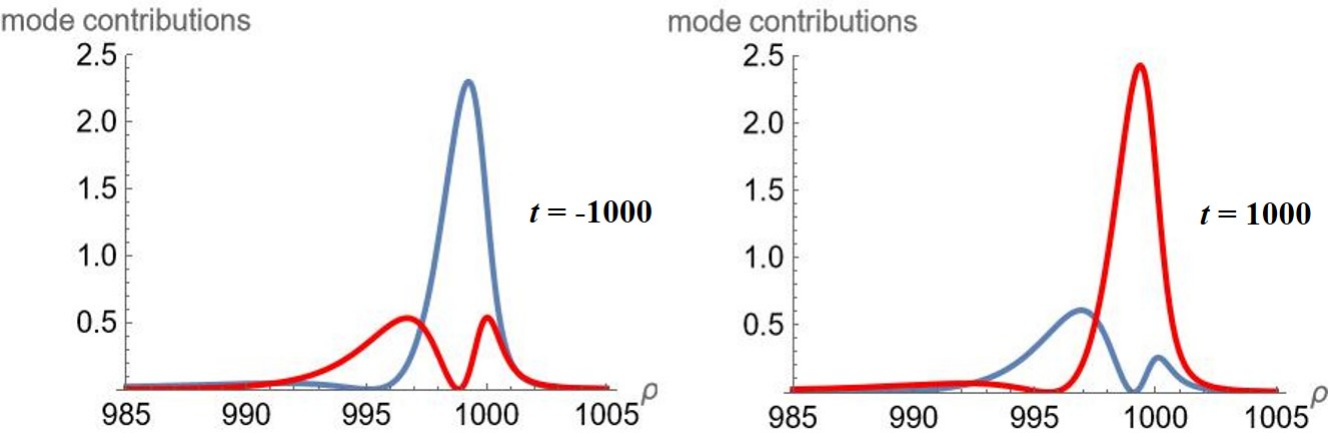}
 \end{minipage}
  \end{tabular}
\caption{
Snapshots at $t=-1000, 1000$ of the gravitational (blue line) and electromagnetic (red line) contributions to 
C-energy density for $(q, p, l) =(-9\sqrt{5}/16, \sqrt{5}/16, 3/4) $
}
\label{fig:fig13}
\end{figure}

\medskip
Therefore, let us first present the corresponding expressions for the gravitational and electromagnetic contributions to the C-energy density, evaluated at past or future null infinity. Similar to Eq.~(\ref{eq:Infoccupancyeachmode(a)(ii)}) and utilizing the functions defined in (\ref{eq:Infdfunctiondef(a)(ii)}), the formula for case~(b) with seed  (ii) can be specified as follows:
\begin{eqnarray}
{}^{(-)}{\cal E}_{\rm g}(v) &=& 4{\cal K}_{+}(v|p,q,l), \ \ \ \ \ 
{}^{(-)}{\cal E}_{\rm em}(v) = 4{\cal K}_{\times}(v|p,q,l),  \nonumber   \\
{}^{(+)}{\cal E}_{\rm g}(u) &=& 4{\cal K}_{+}(-u|p,q,-l),\ \ \ \ \ 
{}^{(+)}{\cal E}_{\rm em}(u) = 4{\cal K}_{\times}(-u|p,q,-l), 
           \label{eq:InfoccupancyG-EM(b)(ii)}
\end{eqnarray}
where the subscripts $\rm g$ and $\rm em$ denote gravitational and electromagnetic modes, respectively. From this, we can anticipate non-trivial conversions between gravitational and electromagnetic modes when comparing past and future null infinities. Indeed, Fig.~\ref{fig:fig11} illustrates an example of such non-trivial conversion. In cases depicted in Figs.~\ref{fig:fig12} and \ref{fig:fig13}, a wave with a dominant gravitational mode incident near the axis results in the electromagnetic modes being significantly amplified by nonlinear effects. Subsequently, after departing from the axis, the wave continues to propagate while maintaining the enhanced electromagnetic mode.

\medskip
Finally, we will discuss the extent to which conversion occurs when waves are observed at past and future null infinities, although a systematic mathematical analysis has not yet been successful. To advance the analysis, let us first introduce the following quantities:
\begin{eqnarray}
{}^{(-)}\gamma_{\rm em/g} := 2\int_{-\infty}^{\infty} {}^{(-)}{\cal E}_{\rm em/g}(v)\ dv, \ \ \ 
{}^{(+)}\gamma_{\rm em/g} := 2\int_{-\infty}^{\infty} {}^{(+)}{\cal E}_{\rm em/g}(u)\ du, 
    \label{eq:Tem}
\end{eqnarray}
where ${}^{(-)}{\cal E}_{\rm em/g}(v)$ and ${}^{(+)}{\cal E}_{\rm em/g}(u)$ have already been given in Eq.~(\ref{eq:InfoccupancyG-EM(b)(ii)}). Doubling the quantities is necessary to ensure that ${}^{(\pm)}\gamma_{\rm g} + {}^{(\pm)}\gamma_{\rm em}$ equals the total C-energy (here $\ln(q/p)^4$).
Then, using these quantities ${}^{(-)}\gamma_{\rm em}$ and ${}^{(+)}\gamma_{\rm em}$, we introduce the ratio of the electromagnetic contribution to the C-energy at future null infinity to the electromagnetic contribution to the C-energy at past null infinity:
\begin{eqnarray}
{\rm Ratio} := \frac{{}^{(+)}\gamma_{\rm em} }{{}^{(-)}\gamma_{\rm em} }.
    \label{eq:Ratio}
\end{eqnarray}

%

\begin{figure}[h]
  \begin{tabular}{c}
\hspace{-10mm} \begin{minipage}[t]{1.0 \hsize}
 \centering
\includegraphics[width=15cm]{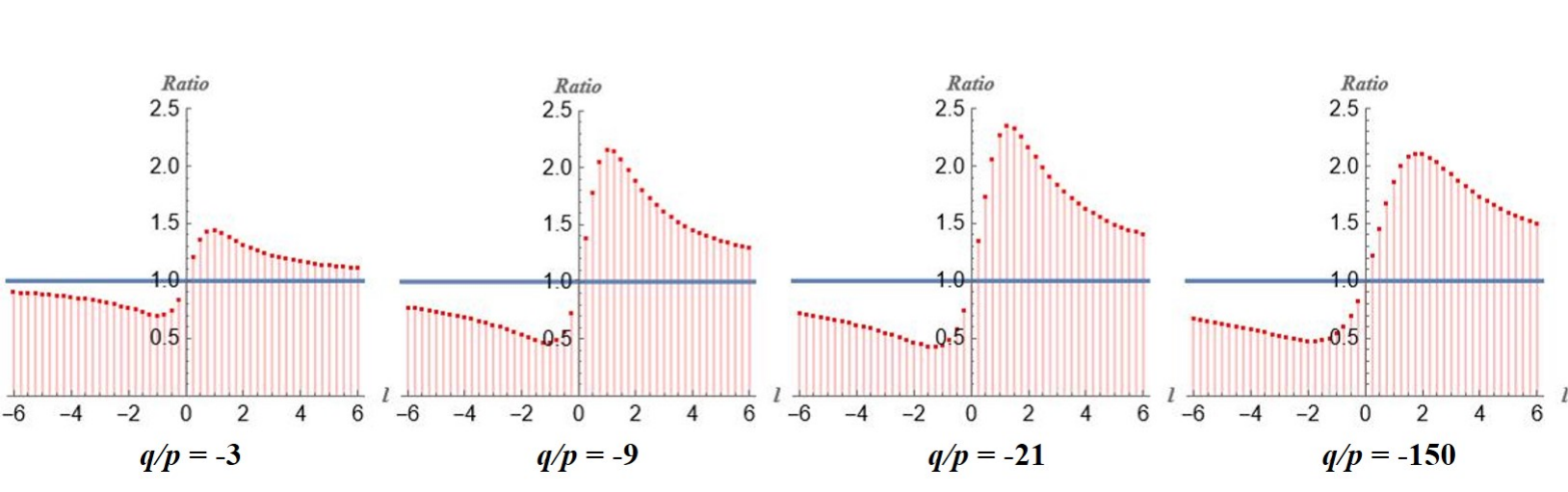}
 \end{minipage}
  \end{tabular}
\caption{
The ratio of the electromagnetic contribution to the C-energy at future null infinity to the electromagnetic contribution to the C-energy at past null infinity: Each figure, from left to right, corresponds to to $q/p=-3,\, -9,\, -21,\, -150$, respectively.  
}
\label{fig:fig14}
\end{figure}

\medskip
To evaluate the integral numerically, the upper and lower limits of the integration range, originally set to $\pm\infty$, are replaced with sufficiently large finite values, $\pm1000$. For example, the ratios corresponding to various C-energies (i.e., $\ln(q/p)^4$) are depicted in Fig.~\ref{fig:fig14}, the horizontal axes of which represent the parameter $l$. 
It is important to note that the parameters must satisfy the constraint equation $q^2-p^2-l^2=1$; given two parameters $l$ and $q/p$, the value of the ratio can be uniquely determined. 
This constraint equation also restricts the range of the parameter $q/p$ to $|q/p| > 1$. A useful fact is that by using the formulas in (\ref{eq:FormulasforK}), we can demonstrate that the ratio corresponding to the case $q/p < 0$ and $l$ is equal to the ratio corresponding to the case $q/p > 0$ and $-l$, thus studying only the former case is sufficient. 
By examining several graphs, including those in Fig.~\ref{fig:fig14}, we can discern qualitative features of the conversion phenomena, albeit within the scope of the phenomena represented by the solutions treated here. 
It is anticipated that upper and lower bounds, which are positive definite, exist for the amplification factor of electromagnetic modes, respectively.
This fact is certainly ensured by the positivity of the function ${\cal K}_{\times}(x|p,q,l)$, except for some points where it becomes zero, as can be deduced from inspection of Eq.~(\ref{eq:Infdfunctiondef(a)(ii)}).
Indeed, from several graphs, including Fig.~\ref{fig:fig14}, we can predict that the upper limit of amplification is approximately $2.3$ and the lower limit of suppression is approximately $0.4$. 
Furthermore, when the parameter $l$ is zero or approaches close to $\pm\infty$, the ratio appears to approach one. 
The latter fact, however, can be easily deduced, due to the fact that the $l$-dependence of the ratio~(\ref{eq:Ratio}) is represented by using Eqs.~(\ref{eq:Infdfunctiondef(a)(ii)}) and (\ref{eq:InfoccupancyG-EM(b)(ii)}), as follows:
\begin{eqnarray}
{\rm Ratio} = \frac{P(q/p)\,l^2 -Q(q/p)\,l +R(q/p) }{P(q/p)\,l^2 +Q(q/p)\,l +R(q/p) },
    \label{eq:Ratio2}
\end{eqnarray}
where $P$, $Q$, and $R$ are constants determined by $q/p$, and also $P$ and $R$ are positive.

\vskip 2cm

\section{Summary and Discussion}

In this study, we have used the composite harmonic mapping method to broaden the Einstein-Maxwell system solution previously examined, aiming to uncover new phenomena driven by the system's nonlinear dynamics. Specifically, we focused on the phenomena of mode conversion between gravitational and electromagnetic waves. Utilizing the exact solutions constructed by the composite mapping method, we analyzed three scenarios: case~(a) with seed~(i), case~(a) with seed~(ii), and case~(b) with seed~(ii). In cases (a) and (b), we utilized the complex line and the totally real Lagrangian plane, respectively, as totally geodesic surfaces within $\rm{H^2_C}$. For both seed~(i) and seed~(ii), we employed harmonic maps constructed using a linear wave function $\tau$ and a solitonic vacuum solution, respectively.

\medskip
In our expanded examination of case~(a) with seed~(i), previously addressed in our prior work, we present a more comprehensive analysis that includes a detailed description of the solution and its intriguing features. Specifically, we have enhanced the occupancy diagram to a two-dimensional format, offering a clearer view that facilitates the identification of significant conversion phenomena between gravitational and electromagnetic modes near the axis.
The updated occupancy diagrams reveal diverse wave behaviors close to the axis, particularly in terms of conversion phenomena. However, any changes in the occupancy ratio between gravitational and electromagnetic modes near the axis invariably revert to their original state as the waves move away from the axis.
This observation prompts an important question: Is there ever a lasting difference in the occupancy of gravitational and electromagnetic modes, even when the incident and reflected waves are distant from the axis? Our research into subsequent cases has provided insights into this query.

\medskip
In both cases (a) and (b) with seed~(ii), solutions were constructed using the solitonic vacuum solution~\cite{Economou}, governed by three parameters $\{q, p, l\}$ that satisfy the constraint equation $q^2-p^2-l^2=1$. These solutions from both cases demonstrate nontrivial and varied mode conversions near the axis. Observing these waves from a distance, case~(a) shows no conversion between gravitational and electromagnetic waves, whereas case~(b), except when $l=0$, exhibits significant conversion. Additionally, as outlined in the previous section, finite upper and lower limits exist for the amplification and attenuation of mode contributions to C-energy at past and future null infinities. Specifically, in case~(b) for seed~(ii), the upper and lower bounds on the amplification factor are approximately 2.4 and 0.4, respectively.

\medskip
From both a fundamental and applied physics perspective, determining the extent of conversion between wave types presents an important and intriguing question. 
If large amplifications occur when gravitational waves, accompanied by minor electromagnetic waves, imploded in a finite region without collapsing into a black hole, such as a sufficiently elongated object,  significant electromagnetic waves could be emitted, amplified by the energy from the gravitational waves. 
Notably, the interaction of gravitational waves is very weak, allowing them to concentrate significantly even if only a small amount of matter is distributed. In contrast, electromagnetic waves, due to their strong interaction, do not behave this way. Thus, it is expected that unwieldy energy in incident waves could be converted into more manageable energy in reflected waves. Although not practically feasible, exploring the extent of this conversion might yield interesting insights.
Within the approach of using the exact solutions, other types of soliton solutions may  provide more interesting and novel conversion phenomena~\cite{Papadopoulos:1990pk,Papadopoulos:1995nj,Tomizawa:2015zva}.

\acknowledgements
We would like to thank Nakao Ken-ichi, Nakamura 
Kouji, Tomohiro Harada and Tsutomu Kobayashi for helpful comments.   
TM was supported by the Grant-in-Aid for Scientific Research (C) [JSPS KAKENHI Grant Number~20K03977], and ST was supported by the Grant-in-Aid for Scientific Research (C) [JSPS KAKENHI Grant Number~21K03560] from the Japan Society for the Promotion of Science.

\appendix
\section{composite harmonic mapping method}\label{appendix:A}
When a map $\varphi$ from a base space $M(x^{\alpha})$, whose metric $h_{\mu\nu}$, to a target space $N(\varphi^{A})$, whose metric $G_{AB}$,  satisfies the following equation: 
\begin{eqnarray}
\nabla^2 \varphi^{A}
 + \Gamma^{A}_{\,BC}\,\,h^{\alpha \beta}\nabla_{\alpha} \varphi^{B} \nabla_{\beta} \varphi^{C}
 =0, 
  \label{eq:harmonic map eq} 
\end{eqnarray}
the map $\varphi$, i.e. $\varphi^{A}(x^{\alpha})$, is called a harmonic map.
Here, $\Gamma^{A}_{\,BC}$ means Christoffel symbol corresponding to  $G_{AB}$.
In general, the harmonic map equation~(\ref{eq:harmonic map eq}) is difficult to solve directly, so that the composite harmonic mapping method have been used widely, as a simple and convenient method.

The essence of the idea is that, to reduce the difficulty, by embedding an appropriate totally geodesic subspace $K(v^{i})$ in  the target space $N$, the map is divided into two steps:
\[
\varphi(x)=\tilde{\varphi}(v(x)) \,: \  M(x^{\alpha}) \longrightarrow \,K(v^{i}(x) )
                                   \longrightarrow \,N(\tilde{\varphi}^{A}(v) ).
\]
As shown in the above diagram, the first map $v$ transforms the base space $M$ into the intermediate space $K$,  
and the second map $\tilde{\varphi}$ embeds $K$ into the final target space $N$ as a totally geodesic subspace. 
Then the composite map $\varphi$ whose explicit expression is $\tilde{\varphi}(v(x))$ can give a map from $M$ to $N$. 
If the map satisfies the harmonic map equation~(\ref{eq:harmonic map eq}), this composite map becomes a harmonic map.
If we substitute the composite map function $\tilde{\varphi}(v(x))$ into Eq.~(\ref{eq:harmonic map eq}), the equation is transformed into the following form: 
\begin{eqnarray}
0 &=&
\left( \nabla^2 v^{k}+ \Gamma^{k}_{\,ij} h^{\alpha\beta} \del{\alpha}v^{i}\del{\beta}v^{j} \right) \del{k}\tilde{\varphi}^{A} 
  \nonumber \\
& &  
 + \left( \del{i}\del{j}\tilde{\varphi}^{A} -\Gamma^{k}_{\,ij}\del{k}\tilde{\varphi}^{A} +\Gamma^{A}_{\,BC}\, \del{i}\tilde{\varphi}^{B} 
 \del{j} \tilde{\varphi}^{C} \right) h^{\alpha\beta} \del{\alpha}v^{i}\del{\beta}v^{j}. 
\end{eqnarray}
Therefore, the following set of equations can be considered a sufficient condition for the 
$\varphi(x)$ to be a harmonic map: 
\begin{eqnarray}
\begin{cases}
0= \nabla^2 v^{k}+ \Gamma^{k}_{\,ij} h^{\alpha\beta} \del{\alpha}v^{i}\del{\beta}v^{j},  \\
0= \del{i}\del{j}\tilde{\varphi}^{A} -\Gamma^{k}_{\,ij}\del{k}\tilde{\varphi}^{A} +\Gamma^{A}_{\,BC}\, \del{i}\tilde{\varphi}^{B} 
 \del{j} \tilde{\varphi}^{C}.
 \end{cases}
  \label{eq:TGC}  
\end{eqnarray}
From the first equation, the map corresponding to $v^{i}(x)$ means a harmonic map  from the base space $M$  to the intermediate space $K$.
On the other hand, the second equation imposes that the submanifold embedded by the map $\tilde{\varphi}^{A}(v)$ is totally geodesic.
The mathematical foundation of the method adopted here is given in~\cite{EellsSampson64}.


\end{document}